\documentclass{aa}  
\usepackage{graphicx}
\usepackage[breaklinks,pdfnewwindow=true,colorlinks=true,citecolor=blue, urlcolor=blue]{hyperref}
\usepackage{comment}
\usepackage{subcaption}
\usepackage{xcolor}
\usepackage{dblfloatfix}
\usepackage{comment}
\usepackage{multirow,bigdelim,dcolumn,booktabs}
\usepackage{mathtools}
\usepackage[mathscr]{eucal}
\usepackage{placeins}
\usepackage{threeparttable}
\usepackage{lscape}

\makeatletter
\newcommand{\vast}{\bBigg@{4.1}}
\newcommand{\Vast}{\bBigg@{5}}
\usepackage{txfonts}
\usepackage{orcidlink}
\newcommand{\orcid}[1]{\unskip\protect\href{https://orcid.org/#1}{\protect\includegraphics[width=8pt,clip]{logo_orcid}}}
\newcommand{\ohhdp}{o-H$_2$D$^+$\:}
\newcommand{\nndp}{N$_2$D$^+$\:}
\newcommand{\nnhp}{N$_2$H$^+$\:}
\newcommand{\Xo}{$X$(o-H$_2$D$^+$)\:}
\newcommand{\Xn}{$X$(N$_2$D$^+$)\:}

\newcommand{\Rd}   {R_{\rm D}}
\newcommand{\fD}   {f_{\rm D}}

\newcommand{\Tdust} {T_{\rm dust}}

\begin{document} 
	
	\title{Time evolution of o-H$_2$D$^+$, N$_2$D$^+$, and N$_2$H$^+$ during the high-mass star formation process}
	\author{G.~Sabatini\inst{1,2\orcidlink{0000-0002-6428-9806}}
		\and
		S.~Bovino\inst{1,3,4}\orcidlink{0000-0003-2814-6688}
		\and
		E. Redaelli\inst{5}\orcidlink{0000-0002-0528-8125}
		\and
		F.~Wyrowski\inst{6}\orcidlink{0000-0003-4516-3981}
		\and
		J.~S.~Urquhart\inst{7}\orcidlink{0000-0002-1605-8050}
		\and 
		A.~Giannetti\inst{8,9}\orcidlink{0000-0003-3869-6501}
		\and
		J.~Brand\inst{2}\orcidlink{0000-0003-1615-9043}
		\and
		K.~M.~Menten\inst{6}\orcidlink{0000-0001-6459-0669}
	}
	
	\institute{
		INAF, Osservatorio Astrofisico di Arcetri, Largo E. Fermi 5, I-50125, Firenze, Italy; \email{giovanni.sabatini@inaf.it}
		\and 
		INAF, Istituto di Radioastronomia - Italian node of the ALMA Regional Centre (It-ARC), Via Gobetti 101, I-40129 Bologna, Italy
		\and
		Chemistry Department, Sapienza University of Rome, P.le A. Moro, 00185 Rome, Italy
		\and 
		Departamento de Astronom\'ia, Facultad Ciencias F\'isicas y Matem\'aticas, Universidad de Concepci\'on, Av. Esteban Iturra s/n Barrio Universitario, Casilla 160, Concepci\'on, Chile
		\and
		Centre for Astrochemical Studies, Max-Planck-Institut f\"ur extraterrestrische Physik, Gie{\ss}enbachstra{\ss}e 1, 85748 Garching bei M\"unchen, Germany
		\and
		Max-Planck-Institut f\"ur Radioastronomie, Auf dem H\"ugel, 69, 53121, Bonn, Germany
		\and
		Centre for Astrophysics and Planetary Science, University of Kent, Canterbury, CT2 7NH, United Kingdom
		\and 
		INAF, Istituto di Radioastronomia, Via Gobetti 101, I-40129 Bologna, Italy
		\and
		Instituto de Radioastronomía y Astrof\'{i}sica UNAM, Apartado Postal 3-72 (Xangari), 58089 Morelia, Michoac\'{a}n, M\'{e}xico
	}
	
	\date{Received:~\today}
	
	
	\abstract
	{Deuterium fractionation is a well-established evolutionary tracer in low-mass star formation, but its applicability to the high-mass regime remains an open question. In this context, the abundances and ratios of different deuterated species have often been proposed as reliable evolutionary indicators for different stages of the high-mass star formation process.} 
	{In this study, we investigate the role of N$_2$H$^+$ and key deuterated molecules (o-H$_2$D$^+$ and N$_2$D$^+$) as tracers of the different stages of the high-mass star formation process. We assess whether their abundance ratios can serve as reliable evolutionary indicators.}
	{We  conducted APEX observations of \ohhdp(1$_{10}$ – 1$_{11}$), \nnhp(4-3), and \nndp(3-2) in a sample of 40 high-mass clumps at different evolutionary stages, selected from the ATLASGAL survey. Molecular column densities and abundances relative to H$_2$, $X$, were derived through spectral line modelling, both under local thermodynamic equilibrium (LTE) and non-LTE conditions.} 
	{The \ohhdp column densities show the smallest deviation from LTE conditions when derived under non-LTE assumptions. In contrast, \nnhp shows the largest discrepancy between the column densities derived from LTE and non-LTE. In all the cases discussed, we found that $X$(o-H$_2$D$^+$) decreases more significantly with each respective evolutionary stage than in the case of $X$(N$_2$D$^+$); whereas $X$(N$_2$H$^+$)  increases slightly. Therefore, the validity of the $X$(o-H$_2$D$^+$)/$X$(N$_2$D$^+$) ratio as a reliable evolutionary indicator, recently proposed as a promising tracer of the different evolutionary stages, was not observed for this sample. While the deuteration fraction derived from N$_2$D$^+$ and N$_2$H$^+$ clearly decreases with clump evolution, the interpretation of this trend is complex, given the different distribution of the two tracers.}
	{Our results suggest that a careful consideration of the observational biases and beam-dilution effects are crucial for an accurate interpretation of the evolution of the deuteration process during the high-mass star formation process.} 

\keywords{Astrochemistry --
	Star: formation --
	ISM: molecules --
	Molecular processes
}
\maketitle

\section{Introduction}
Massive stars (i.e. $M_\star > 8$ M$_\odot$) are crucial for the energy balance of their host galaxies. They shape the properties of their environments at the different galactic scales through radiation, winds and supernova explosions \citep[e.g.][]{Kennicutt05}, and enrich the interstellar medium (ISM) with heavy elements \citep[e.g.][]{Smartt09}. The latter pave the way for the formation of complex organic molecules (COMs; namely, organic molecules with at least six atoms; \citealt{Herbst-vanDishoeck09, Ceccarelli23}). However, due to the difficulties in observing high-mass star-forming regions (HMSFRs) at different stages of their life cycle \citep[e.g.][]{Zinnecker&Yorke07}, the formation and evolution of massive stars is still poorly understood. One of the major challenges is to determine the ages of massive young stellar objects, which depends on a wide range of physical conditions and evolutionary phases, from cold and quiescent infrared dark clouds (IRDCs) to hot and ionised H{\small II} regions.\\
\indent One way to estimate the age of HMSFRs is to use chemical clocks, namely, molecular species whose abundances change significantly over time due to chemical reactions, thermal processes, or selective depletion and desorption on dust grains (e.g. \citealt{Fontani07, Beuther09, Hoq13, Bovino19,Bovino21, Giannetti19, Urquhart19, Sabatini20, Sabatini21}; see also \citealt{vanDishoeck98} for a review). Among the various chemical clocks proposed in the literature, deuterated species, such as N$_2$D$^+$, and ortho-H$_2$D$^+$ (hereafter, o-H$_2$D$^+$) have received particular attention because of their strong response to gas temperature and ionisation fraction, both of which are expected to increase as star formation progresses (e.g. \citealt{BerginTafalla07} and \citealt{Socci24}). Furthermore,
$\rm H_2D^+$ forms from the trihydrogen cation $\rm H_3^+$ reacting with deuterated molecular hydrogen HD, a reaction that is particularly efficient when the temperature is low and the H$_2$ ortho-to-para ratio is low \citep{Dalgarno84, Pagani92_model}. Then, \nnhp and \nndp are formed from $\rm N_2$, a late-type molecule that can take about ten times longer than CO to form \citep[cf.][and references therein]{HilyBlant10}. These species are predominantly formed and destroyed via gas-phase reactions. They have two main destruction routes: dissociative recombination with free electrons or reactions with carbon monoxide (e.g. \citealt{Loison19, Redaelli20, Oberg21}, and references therein).\\
\indent However, even though N$_2$D$^+$ is an abundant ion in cold and dense gas, it has been recently established that it is affected by depletion at high densities (i.e. $n$(H$_2$)$\gtrsim$10$^5$ cm$^{-3}$; \citealt{Redaelli19}). Therefore, it is not an ideal tracer of the prestellar stages. In addition, based on the anti-correlated distribution between o-H$_2$D$^+$ and N$_2$D$^+$ reported by \cite{Giannetti19}, it was recently shown that relative to o-H$_2$D$^+$, which forms very early on and thus exclusively traces cold, prestellar gas, the (3-2) transition of N$_2$D$^+$ (with a critical density, $n_\mathrm{cr}>$ 10$^{5.5}$ cm$^{-3}$; see \citealt{Li22b}) appears at relatively later stages of the star-formation process surviving up to the early phase of protostellar activity.\\
\indent Attempts to confirm this anti-correlation was pursued by \cite{Miettinen20}, using Atacama Pathfinder Experiment (APEX) observations in three prestellar and three protostellar cores in the Orion-B9 filament. This work confirmed the downward trend for o-H$_2$D$^+$ with evolution, while reporting a large spread with an unclear behaviour for N$_2$D$^+$. However, the authors justified this result with a spatial offset between the two tracers.\\
\indent High-resolution interferometric observations \citep{Kong16} have shown that cores in evolved stages have higher N$_2$D$^+$ abundances compared to o-H$_2$D$^+$, confirming the fact that N$_2$D$^+$ is forming at later stages and under different physical conditions. Similar results were also reported in \cite{Li22b}, who find a detection rate of N$_2$D$^+$ in protostellar environment higher than the one obtained in prestellar cores. \cite{Redaelli22} reported well-correlated column
densities for the two tracers by identifying cores in o-H$_2$D$^+$ in the 70~$\mu$m dark clump G014.492-00.139 \citep[e.g.][]{Sanhueza19, Morii23, Izumi24}. However, this result is not in contradiction with \cite{Giannetti19}, who analysed cores identified in sub-millimeter/far-infrared (sub-mm/FIR) wavelength continuum emission. This might indeed be insufficient if we are trying to distinguish between pre- and proto-stellar stages and can lead to unreliable results in the identification of prestellar phases; therefore, this may also be an obstacle when assessing the evolutionary behaviour of chemical clocks \citep{Redaelli21}. Finally, when comparing single dish and interferometric results, we have to consider that the prior look at the average properties of the clumps, which smears out the chemical response to physical changes occurring at the smaller scales mapped by high-resolution observations \citep[see][]{Sabatini22}.\\
\indent In this work, we report a comprehensive study of these molecules in HMSFRs, with the aim to further explore the chemical connection between deuterated species in a larger sample of independent clumps at distinct evolutionary stages. The paper is organised as follows: In Sect.~\ref{sec2:sample}, we describe the observational sample, Sects.~\ref{sec3:data} and~\ref{sec4:results}, are focused on the data reduction and the data analysis, respectively. In Sect.~\ref{sec5:dicussion}, we discuss the result within the context of HMSFRs and make a comparison with previous works.  

\section{The sample}\label{sec2:sample}
The APEX {Telescope Large Area Survey of the Galaxy} (ATLASGAL\footnote{\url{https://atlasgal.mpifr-bonn.mpg.de/cgi-bin/ATLASGAL_DATABASE.cgi}}; \citealt{Schuller09}) provides a solid basis for a complete characterisation of a large number of massive star forming clumps in all their evolutionary phases (e.g. \citealt{Molinari08}). The ATLASGAL {Compact Source Catalog} contains about $10^4$ clumps identified in their $870~\mu$m continuum emission (e.g. \citealt{Contreras13}, \citealt{Urquhart14b}). Together with follow-up spectral line observations, it provides reliable estimates of kinematic distances and location in the Galactic plane, total masses, bolometric luminosities, dust temperature, and age distributions \citep[][and \citealt{Sabatini21}]{Urquhart14c, Wienen15, Urquhart18}.

\subsection{The Top100 sample}

\indent In the ATLASGAL framework, the TOP100-sample (hereafter, TOP100; see \citealt{Giannetti14}) has been defined as a flux-limited sample of 111 dense/massive clumps, selected with additional infrared (IR) criteria to include sources potentially covering the whole spectrum of ages \citep[see][]{Konig17}. Using the APEX-12m, Mopra-22m and the IRAM-30m single-dish telescopes, a comprehensive spectral survey, covering more than 120~GHz of bandwidth\footnote{Not yet fully published.}, was performed in the frequency range from 80 to 345~GHz for all the TOP100 sources. This survey includes three spectral windows and contains a variety of molecular lines, including COMs. The gas excitation parameters and molecular column densities were derived by carefully analysing a subset of chemical species (i.e., C$^{17}$O, C$^{18}$O, $^{13}$CO, $^{13}$C$^{18}$O, \citealt{Giannetti14}; SiO and $^{29}$SiO, \citealt{Csengeri16}; CH$_3$CN, CH$_3$CCH, and CH$_3$OH, \citealt{Giannetti17_june}; \mbox{ortho- and para-H$_2$CO}, \citealt{Tang18}).
From this survey, \cite{Giannetti17_june} showed a correlation of several quantities (temperature, column densities of different chemical tracers, and H$_2$ volume density) with the evolutionary class and the luminosity-to-mass ratio ($L/M$; a robust evolutionary indicator of the star formation processes, e.g. \citealt{Saraceno96, Molinari08}), highlighting a pattern from the initial contraction phase of the clumpy material to the gradual accretion onto nascent young stellar objects (YSOs). Consequently, the evolutionary sequence outlined in the TOP100 holds statistical significance and, within this context, this sample offers an unbiased and comprehensive subset of massive clumps that reproduce the properties of the entire clump population in ATLASGAL. 


\subsection{Sub-sample selection}\label{sec2.2:subsample}
We select a sub-sample of the TOP100 by excluding the 22 H{\small II} regions, as they are associated with the lower abundances of deuterated molecules \citep[e.g.][]{Fontani11, Urquhart19, Sabatini20}. For the remaining sources, we add a second selection criterion that considers only the sources with a heliocentric distance $<$4~kpc to mitigate beam-dilution effects. This leads to a final sub-sample of 40 massive star-forming clumps (see Appendix~\ref{sec:appB}). We note that the range of observed and inferred physical properties in this sub-sample in terms of total clump mass, bolometric luminosity, column and volume density, and temperature are consistent with the average properties of the different evolutionary classes defined in ATLASGAL and recently updated by \cite{Urquhart22}. The classification was based on a careful analysis of the available multi-wavelength data of the GLIMPSE and MIPSGAL surveys (3-24 $\mu$m; \citealt{Churchwell09} and \citealt{Carey09}), complemented by the additional information of the 70~$\mu$m maps and the HiGAL catalogue \citep{Molinari10, Elia21}.\\
\indent In most cases, the resulting classification is similar to that originally presented by \cite{Konig17}, and comprises four evolutionary stages: (1) quiescent: constituted by sources clear of any large-scale extended 70~$\mu$m emission, devoid of embedded objects and therefore cold ($T_{\rm dust}$$\sim$10-15~K) clumps with $L/M$~$\lesssim$~1.5. This stage is associated with the earliest phase of massive star formation and mainly comprises starless or prestellar cores. Around 13\% of the sources in our sample (five in total) belong to this class; (2) protostellar: composed of warmer clumps where at least one compact source is detected at 70 $\mu$m (but still undetected at 24 $\mu$m), within 12\arcsec~of the centre of the clump, and associated with young clumps likely dominated by cold gas with 1.5~$\lesssim~L/M$~$\lesssim$~5. Eight sources are classified as protostellar in our sample; (3) YSOs: as a result of the progressive heating of the dust driven by the forming (proto-)stars, this stage includes clumps showing stronger emission in the mid-infrared at 3-8 and 24 $\mu$m, and 5~$\lesssim~L/M$~$\lesssim$~15. In this study, 18 sources are associated with this evolutionary stage; (4) H{\small II} regions: this is the most evolved stage of the sequence, consisting of clumps in which massive stars have formed almost entirely. These clumps are bright at 3-70~$\mu$m, with clear detection at 5~GHz. On average, the $L/M$ of the sources associated with this class is $\gtrsim$~15. As mentioned above, none of the sources in our sample were associated with this evolutionary phase. An additional fifth category labelled "ambiguous", is used to classify sources that are not included in the above groups. This category does not correspond to a specific evolutionary stage and is primarily composed of complexes in which several sub-regions at different evolutionary stages are close to each other or of clusters associated with photon-dominated regions (PDR) at the edge of evolved H{\small II} regions. Our final sample includes nine sources in this category.\\
\indent All the evolutionary parameters evolve consistently with the original classification and show only very minor deviations compared to \cite{Giannetti17_june}. These parameters are shown in Table~\ref{tab:comparison}. During the transition from the quiescent phase to the advanced evolutionary phases associated with the presence of a YSO,  average dust temperature, and H$_2$ column density increase from 16 to 23~K and from $\sim$5.0$\times 10^{22}$ to $\sim$1.2$\times 10^{23}$~cm$^{-2}$, while the total mass and luminosity of the clumps increase from $\sim$10$^2$ to $\sim$2.0$\times 10^3$~M$_\odot$ and from $\sim$10$^2$ to $\sim$4.6$\times 10^4$~L$_\odot$, respectively, resulting in an increase of the average $L/M$ from 3 to 33.

\section{Observations and data reduction}\label{sec3:data}
In this work, we investigate how the abundances of multiple molecular tracers (i.e. N$_2$D$^+$, \nnhp and o-H$_2$D$^+$) vary across the different evolutionary stages of ATLASGAL clumps. The APEX datasets employed in this analysis are described in the following subsections.

\subsection{{\rm N$_2$D$^+$} (deuterated diazenylium)}\label{sec3.1:n2dp}
We observed the \nndp(3-2) spectra with the nFLASH dual-frequency MPIfR-PI receiver, an evolved version of the original First Light APEX Submillimeter Heterodyne receiver (FLASH, \citealt{Klein14}) mounted on the APEX telescope (\citealt{Gusten06}), in the on-off (ONOFF) observing mode, assuming a rest frequency of 231.32167~GHz \citep[][see Table~\ref{tab:observations}]{Amano05b}.
The receiver allows simultaneous observations of two 4~GHz sidebands recorded by two fast Fourier transform spectrometer (FFTS) processor units that overlap in the middle by about 100~MHz, providing a full coverage of 7.9~GHz. The observations were performed from December 2020 to July 2021 in two projects (IDs: C-0105.F-9715C-2020 and C-0107.F-9711-2021; PIs: S. Bovino and G. Sabatini, respectively), with a native spectral resolution of 0.078 km s$^{-1}$.\\
\indent The final mean system temperatures, $T_{sys}$, ranged between $\sim$120 and $250$ K, depending on the weather conditions. All observations were made with a precipitable water vapour $<$~2.0~mm. At $\sim$231 GHz, the APEX telescope has an effective FWHM beam size of 26\farcs2, with a corresponding main-beam efficiency of $\sim$0.7\footnote{\url{http://www.apex-telescope.org/telescope/efficiency/?yearBy=2021}}. The main-beam efficiency was applied to the spectra after setting the reference frequency to that of \nndp(3-2). We subtracted a first-order polynomial baseline from each spectrum around the masked line. In some rare cases where a first-order baseline was not sufficient, a second- (or third-) order baseline was used. Finally, all the  spectra of each source were averaged to obtain one spectrum per source. The spectral resolution, $\delta\nu$, was smoothed to 0.5 km s$^{-1}$ to improve the signal-to-noise ratio (S/N). Observations were reduced using a GILDAS/CLASS\footnote{\url{https://www.iram.fr/IRAMFR/GILDAS/}} Python interface pipeline.

\begin{table}
	\caption{Observations and molecular properties.}\label{tab:observations}
	\setlength{\tabcolsep}{1.5pt}
	\renewcommand{\arraystretch}{1}
	\centering
	\begin{tabular}{lc|ccc}
		\hline\hline
		\multicolumn{2}{l|}{Molecules}                                  & N$_2$D$^+$ &       \ohhdp          &  N$_2$H$^+$\\
		\hline
		\multicolumn{2}{l|}{Quantum numbers}                              &   (3-2)   & (1$_{10}$ - 1$_{11})$ &     (4-3)  \\
		\multicolumn{2}{l|}{Rest frequencies (MHz)\tablefootmark{a}}                         & 231321.67 & 372421.39 & 372672.51 \\
		\multicolumn{2}{l|}{Beam size ($\arcsec$)}                        &     26    &          17           &     17     \\
		\multicolumn{2}{l|}{$\delta\nu$ (km s$^{-1}$)}              &    0.5    &         0.5           &     0.5    \\
		\multicolumn{2}{l|}{$\langle$rms$\rangle$~(K)\tablefootmark{b}}      &    0.02   &       0.03            &     0.06   \\
		\multicolumn{2}{l|}{References\tablefootmark{c}} &    [1]    &         [2]           &     [1]    \\
		\hline
		\multicolumn{2}{l|}{ log$_{10}$[$n_\mathrm{cr}$~(cm$^{-3}$)]\tablefootmark{d}} & 5.5 & 5.0 & 5.9 \\  
		\multicolumn{2}{l|}{$E_{\rm up}$ (K)\tablefootmark{e}}                    &  22.2 &  17.9 &  44.7 \\  
		\multicolumn{2}{l|}{log$_{10}[A_{\rm ul}$~(s$^{-1}$)]\tablefootmark{e}}   & -3.36 & -3.96 & -2.51 \\  
		\multicolumn{2}{l|}{$g_{\rm u}$\tablefootmark{e}}                         &   63  &   9   &   81  \\  
		\hline
		\multirow{4}{*}{Detection rates $\vast\{$} & Quiescent    & 100\% & 80\% & 100\%\\  
		& Protostellar & 88\%  & 13\% & 100\%\\  
		& YSO          & 67\%  & 10\% & 100\%\\  
		& PDR          & 56\%  & 22\% & 100\%\\   
		
		\hline  
	\end{tabular}
	\tablefoot{\tablefoottext{a}{The rest frequencies are taken from \cite{Amano05} for o-H$_2$D$^+$, and from \cite{Amano05b} for \nnhp and N$_2$D$^+$.} \tablefoottext{b}{Temperatures are reported on the main-beam temperature scale}. \tablefoottext{c}{[1]~this work; [2]~\cite{Sabatini20}}. \tablefoottext{d}{Critical densities computed for a temperature $<$20~K, considering downward collision rates in a multilevel system. The collisional rates are taken from \cite{Hugo09} for o-H$_2$D$^+$, \cite{Balanca20} for N$_2$H$^+$ and \cite{Lin20} for N$_2$D$^+$.} \tablefoottext{e}{$g_{\rm u}$ is the statistical weight. Spectroscopic data from The Cologne Database for Molecular Spectroscopy (\citealt{Muller05}). Note: the upper level energy of the \ohhdp is defined here with respect to \ohhdp~1$_{11}$ level (see \citealt{Vastel06})}.}
\end{table}  

\begin{figure*}
	\centering
	{\includegraphics[width=0.96\hsize]{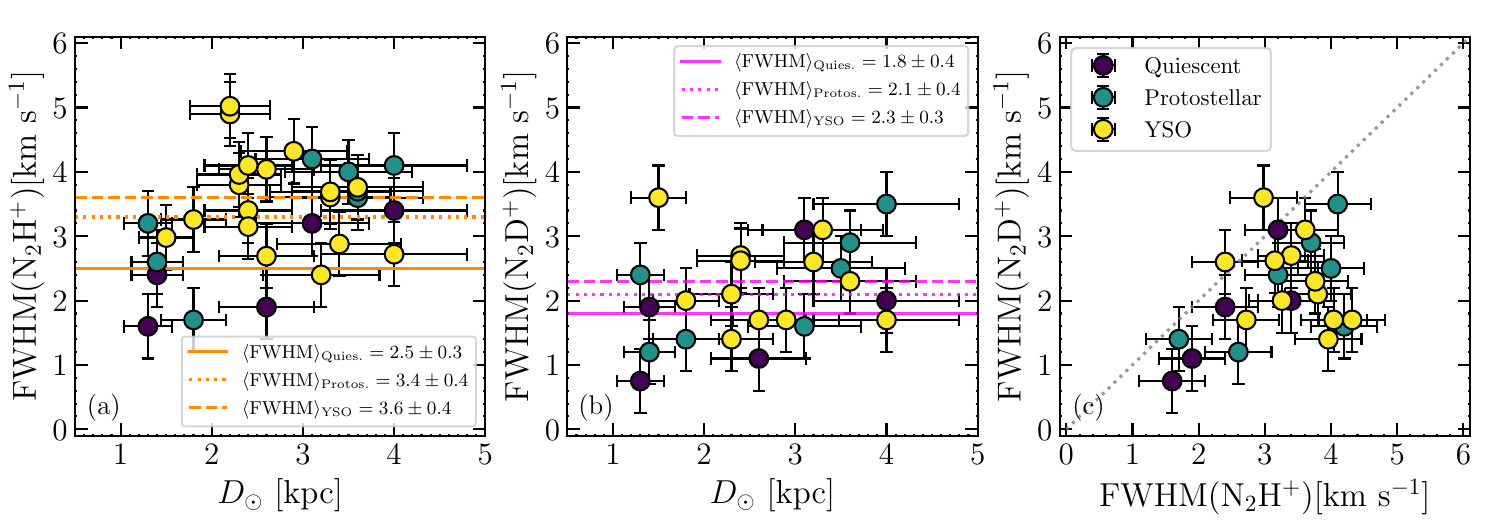}}
	\caption{Heliocentric distance versus the FWHM of  (a) N$_2$H$^+$ and (b) N$_2$D$^+$. Orange and magenta lines represent the median FWHM values of N$_2$H$^+$ and N$_2$H$^+$, respectively, within the quiescent (solid), the protostellar (dotted) and the YSO (dashed) groups. Panel (c) compares the FWHMs of the two molecular tracers, with colours indicating the different evolutionary stages (dark blue: quiescent; green: protostellar; yellow: YSO). The uncertainties are shown as black bars.}\label{fig:FWHM}
\end{figure*}

\subsection{Additional tracers: {\rm N$_2$H$^+$} (diazenylium) and {\rm \ohhdp} (deuterated trihydrogen cation) }\label{sec3.2:add_tracers}

The \ohhdp $J_{\rm {K_a, K_c}} = 1_{10} - 1_{11}$ spectra ($\nu_{\rm rest}\sim$372.42139~GHz; \citealt{Amano05}) are part of three APEX projects (IDs: 0101.F-9517; M-097.F-0039-2016 and M-098.F-0013-2016; PI: F. Wyrowski), observed from July 2017 to December 2018. These data were published in \cite{Sabatini20} and we refer to this work for a complete description of the observations and data reduction. 

From the same APEX projects, we have collected further observations of N$_2$H$^+$(4-3) at a rest frequency of $\sim$372.67251~GHz \citep{Amano05b}, which we employed to estimate the deuteration fraction (see Sec.~\ref{sec4:results}). The N$_2$H$^+$(4-3) data are available for the whole sample observed in \nndp(3-2), with a native spectral resolution of 0.038 km s$^{-1}$. The final mean $T_{sys}$ lies between $\sim$600 and 1200 \citep{Sabatini20}. At $\sim$372~GHz, the APEX telescope has an effective FWHM beam size of $\sim$17\farcs5, with a corresponding main beam efficiency of $\sim$ 0.73\footnote{\url{https://www.apex-telescope.org/telescope/efficiency/?yearBy=2017}}. The observations were reduced using GILDAS/CLASS. All the molecular and spectral parameters are summarised in Table~\ref{tab:observations}.

\section{Analysis and results}\label{sec4:results}

Figure~\ref{fig:spectra} presents the observed spectra of \nndp and N$_2$H$^+$, while the spectra of \ohhdp are published in \cite{Sabatini20}. The \nnhp(4-3) line is detected with S/N$>$3 in the whole sample (i.e. detection rate of 100\% over all evolutionary stages defined in ATLASGAL; see Table~\ref{tab:observations}). Conversely, the detection rate of N$_2$D$^{+}$, similar to \mbox{o-H$_2$D$^+$}, decreases with the evolution of the clump (e.g. \citealt{Miettinen20, Sabatini20}), passing from 100\%$\pm$22\% in quiescent sources to 88\%$\pm$18\% and 67\%$\pm$12\% in protostellar and YSO stages, respectively, while slightly more than half of the PDRs (56\%$\pm$17\%) show \nndp detection. The uncertainties in the detection rates are calculated using binomial statistics.

\subsection{Spectral fitting}
To infer the spectral line parameters of \ohhdp, \nndp, and N$_2$H$^+$, we employed {\verb~MCWeeds~} \citep{Giannetti17_june}: a flexible interface between PyMC \citep{Patil10}, a Python library for building Bayesian statistical models, and {\verb~Weeds~} \citep{Maret11}, an extension of the CLASS software\footnote{\url{http://www.iram.fr/IRAMFR/GILDAS}} designed for modelling observed spectra under the assumption of local thermodynamic equilibrium (LTE). We used a fitting algorithm based on Markov chain Monte Carlo (MCMC) techniques to sample the parameter space with non-informative flat priors over the free parameters of the models, i.e. the line central velocity ($V_{\rm lsr}$) and linewidth (FWHM), the molecular column density ($N^{\rm LTE}$), the excitation temperature ($T_{\rm ex}$), and the source size ($\theta_{\rm src}$). For all tracers, we have assumed extended emission with respect to the APEX beam, as their emission scales are usually comparable to the typical clump sizes \citep[e.g.][]{Pillai12, Feng16, Redaelli22, Sabatini23}. The APEX beam sizes at all the frequencies here analysed are smaller than the apparent diameters of the clump as measured by the ATLASGAL survey. However, our observations could be affected by beam dilution at different levels for the distinct tracers and/or evolutionary stages. This is likely not the case for N$_2$H$^+$, as its emission is usually found spread across the whole clump when maps are available \citep[see, for instance][]{Feng16, Redaelli22}, whilst the problem might be more severe for the rarer deuterated species. \cite{Redaelli21, Redaelli22} reported extended \ohhdp emission even in clumps with embedded protostars, suggesting that in the high-density environments typical of high-mass clumps, this molecule is abundant. Concerning N$_2$D$^+$, \cite{Redaelli22} and \cite{Sakai22} reported widespread, resolved emission and strong line emission even in evolved sources, respectively. However, we cannot exclude the possibility that the beam-filling factor is $\eta_{\rm ff} < 1$. We stress that we cannot quantify $\eta_{\rm ff}$ solely on the basis of the data we have at hand, as this would require access to resolved maps of molecular emission. If $\eta_{\rm ff} < 1$, the column densities (and abundances) of \nndp and \ohhdp could be underestimated. However, due to the chemical link between them that we further discuss in Sect.\ref{sec5:dicussion}, we can expect a similar extension for the two species, and even in the case of $\eta_{\rm ff} < 1$, a comparable underestimation of the final column densities.\\
\indent In quiescent sources, the \nnhp lines are detected as a single Gaussian-like component with an average FWHM of $\approx$2.5~km~s$^{-1}$ (see Table~\ref{tab:summary} and Fig.~\ref{fig:FWHM}a). As the sources evolve, the \nnhp lines become broader (FWHM$>$3.6 km s$^{-1}$) and may exhibit multiple velocity components (e.g. 351.45+0.66, G317.87-0.15 and G353.41-0.36).
Conversely, the spectra of \nndp are well fitted by a single velocity component with smaller FWHMs throughout the whole evolutionary sequence (the only exception is the PDR G353.41-0.36). The FWHMs of \nndp lines display less variation across evolutionary stages than \nnhp lines (see Fig.~\ref{fig:FWHM}b). Neither the FWHMs of \nnhp nor \nndp show a significant correlation with the heliocentric distance ($D_{\odot}$; see Table~\ref{tab:summary}) of the sources at the different stages (Fig. \ref{fig:FWHM}a,b), ruling out any distance-related bias in the FWHMs. Instead, the different behaviour found in the class-averaged FWHMs suggest that \nnhp and \nndp probe different regions within the same clumps. \nnhp likely traces larger-scale, more turbulent structures \citep[e.g.][]{Emprechtinger09, Miettinen11, Redaelli22}, while the formation of \nndp is favoured in the denser, colder cores embedded in these structures \citep[e.g.][]{Oberg21}. The narrower \nndp lines reflect this origin (see Fig.~\ref{fig:FWHM}c).

\subsection{Column densities and deuteration fraction}\label{sec:coldens}
\subsubsection{LTE analysis\label{subsec:LTEcol}}
The molecular column densities were determined with {\verb~MCWeeds~}. The best-fit results are shown in Table~\ref{tab:summary} and were
derived assuming an excitation temperature $T_{\rm ex}$ = $T_{\rm dust}$ \citep[e.g.][]{Giannetti19, Sabatini20}. Although the gas-dust thermal coupling condition is virtually always valid under the typical conditions prevalent in high-mass clumps (i.e. when $n$(H$_2$)> 10$^{4.5}$~cm$^{-3}$; \citealt{Goldsmith01}), the H$_2$ volume densities we derive for each clump (assuming a constant H$_2$ distribution and using the clump size) are slightly below the critical densities of the molecular tracers. These derived densities, which represent lower limits, are presented in Table~\ref{tab:summary}, and the critical densities are listed in Table~\ref{tab:observations}. However, we also found that varying the assumed temperature by 40\% affects the column densities within the uncertainties given in Table~\ref{tab:summary}, so that this assumption has only a minor impact on the final results (see also \citealt{Caselli08} and \citealt{Sabatini20}). Furthermore, the assumed $T_{\rm dust}$ values are broadly consistent with gas temperature estimates derived from common tracers like CO and CH$_3$OH at the clump scale for the TOP100 sample \citep{Giannetti17_june}. We also identified and corrected a typo in the version of {\verb~MCWeeds~} used in \cite{Sabatini20}, which led to an overestimation of $N^{\rm LTE}$(o-H$_2$D$^+$) by a factor of $\sqrt{\pi}$ \citep{Redaelli22}. The corrected column densities range from (0.8$\pm$0.5 to 5.0$\pm$1.2)$\times$10$^{12}$~cm$^{-2}$ for o-H$_2$D$^+$, (0.6$\pm$0.1 to 36$\pm$5)$\times$10$^{12}$~cm$^{-2}$ for N$_2$H$^+$, and (0.5$\pm$0.3 to 8.0$\pm$0.9)$\times$10$^{11}$~cm$^{-2}$ for N$_2$D$^+$ (see Table~\ref{tab:summary}). These values are consistent with typical abundances relative to H$_2$ reported in similar environments (e.g. \citealt{Miettinen20, Sabatini20, Li22b}), and are lower than those found by \cite{Crapsi05} in nearby low-mass star-forming regions observed with the IRAM-30m telescope; this is likely due to beam dilution from the larger APEX beam (see also \citealt{Redaelli22} for further discussions on G014.492-00.139). The new $N^{\rm LTE}$(o-H$_2$D$^+$) values vary by 25-50\% compared to \cite{Sabatini20} due to the different temperatures assumed \citep[][see Table~\ref{tab:comparison}]{Urquhart22} and the correction introduced in {\verb~MCWeeds~}.

\subsubsection{Uncertainties due to line opacity and non-LTE conditions}\label{subsec:nonLTEcol}
In this section, we discuss the uncertainties on the $N^{\rm LTE}$ (determined in Sect.~\ref{subsec:LTEcol}) that are due to the assumption of optically thin lines and of LTE-conditions. Both assumptions can lead to an underestimation of the true column densities.\\
\indent Concerning line opacities, \cite{Sabatini20} report optically thin emission for all \ohhdp detections. We verified this assumption with the RADEX code \citep{VanDerTak07} to test the reliability of the {\verb~MCWeeds~} results. As with the column density calculations, we assumed $\eta_{\rm ff} = 1$ when deriving the line opacities. We used the column densities and FWHM values obtained from {\verb~MCWeeds~} and a constant kinetic temperature, $T_{\rm kin}$, equal to $\Tdust$ as input for RADEX. We also assumed that the emission of the different tracers comes from the denser regions of the clump, and for this we set the H$_2$ volume density, $n$(H$_2$), well above the $n_{\rm cr}$ of the individual tracers. We found optical depths, $\tau<0.05$, implying optically thin regimes for all the considered lines.\\
\indent To test the LTE-assumption, we also performed a grid of non-LTE (NLTE) RADEX models, assuming the FWHM derived with {\verb~MCWeeds~} and $T_{\rm kin}=\Tdust$ as input to reproduce the observed lines, leaving the column density and line opacity ($N^{\rm NLTE}$ and $\tau^{\rm NLTE}$) as free parameters. The collisional rates of N$_2$H$^+$, o-H$_2$D$^+$ and N$_2$D$^+$ are taken from  \cite{Balanca20}, \cite{Hugo09} and \cite{Lin20}, respectively. In NLTE analysis, a significant source of uncertainty arises from the assumption of a reliable value of $n$(H$_2$) to reproduce the average environmental properties. This uncertainty is due to the non-uniform gas distribution within the clumps, as different molecular tracers may be associated with distinct regions characterised by different (and variable) densities. To account for this, we considered a range of $n$(H$_2$) values (see Table~\ref{tab:summary}). The lower limit of this range was derived from $N$(H$_2$), assuming a uniform distribution of H$_2$ within each clump and using the effective radius of the clump, $R_{\rm eff}$, which provides a lower limit to the actual density. For the upper limit, we considered a typical uncertainty of a factor of 5 on $R_{\rm eff}$ \citep{Urquhart18, Urquhart22}.\\
\indent Table~\ref{tab:summary} summarises the ranges of assumed $n$(H$_2$), and of the $N^{\rm NLTE}$ and $\tau^{\rm NLTE}$ obtained with the RADEX code in the NLTE case. The $N^{\rm NLTE}$ derived for the \ohhdp line shows the smallest deviation from the LTE results. RADEX models at the lower $n$(H$_2$) yield $N^{\rm NLTE}$ that are 1.2 to 1.6 times higher than $N^{\rm LTE}$. This minor discrepancy is consistent with the fact that the gas density of the clumps is in almost all cases larger than the line $n_{\rm cr}$ (see Table~\ref{tab:observations}). The $\tau^{\rm NLTE}$ are low ($<0.06$) and the  estimated $T_{\rm ex}$ range from 8.6 to 15.5 K, in agreement with the $T_{\rm kin}$ adopted in the LTE case, supporting for \ohhdp ($1_{10} - 1_{11}$) the LTE assumption made in Sect.~\ref{subsec:LTEcol}. We emphasise that the difference between $N^{\rm NLTE}$ and $N^{\rm LTE}$ is even less pronounced at the higher $n$(H$_2$), where $N^{\rm NLTE}$ turns out to be 25\% higher than $N^{\rm LTE}$ at most. A similar result is found for N$_2$D$^+$, for which the estimated $T_{\rm ex}$ range from 5.0 to 18.8 K. The $N^{\rm NLTE}$ overestimates $N^{\rm LTE}$ by factors of 1.3 to 5 at low $n$(H$_2$) and by a factor of 1 to 1.6 at high $n$(H$_2$). The \nndp emission proves to be optically thin over the entire density range explored ($\tau^{\rm NLTE} < 0.16$).\\
\indent Due to its higher $n_{\rm cr}$, the $N^{\rm NLTE}$ derived from \nnhp(4-3) shows the most significant deviations from LTE. At the lowest $n$(H$_2$), the $N^{\rm NLTE}$ differ from the LTE results by factors ranging from 5 to 32. This discrepancy is often associated with optically thick emission ($0.2\lesssim\tau^{\rm NLTE}\lesssim11$). At higher densities, the $N^{\rm NLTE}$/$N^{\rm LTE}$ is reduced to a factor of 1 to 6, and the emission becomes mostly optically thin ($0.1\lesssim\tau^{\rm NLTE}\lesssim2$). The associated $T_{\rm ex}$ range from 5.5 to 15.6 K and, as observed for the other transitions, tend to increase with the evolutionary stage of clumps.\\
\indent It is worth stressing that a NLTE analysis introduces additional uncertainties beyond those of an LTE approach. While both methods require knowledge of the gas temperature, the NLTE analyses also crucially depend on the assumed gas density (see Table~\ref{tab:summary}). This emphasises the larger uncertainty associated with NLTE results, especially when derived based on a single molecular transition. Despite the fact that the column density of \nnhp may be up to a factor of $\sim$30 higher in the NLTE case, we anticipate here that the evolutionary trends of the abundances and abundance ratios given in the following sections are not affected by the choice of analysis. In the following, we show the results obtained with the LTE-analysis and discuss the differences with the NLTE case that are relevant for our conclusions.

\begin{figure}
	\centering
	{\includegraphics[width=1\columnwidth]{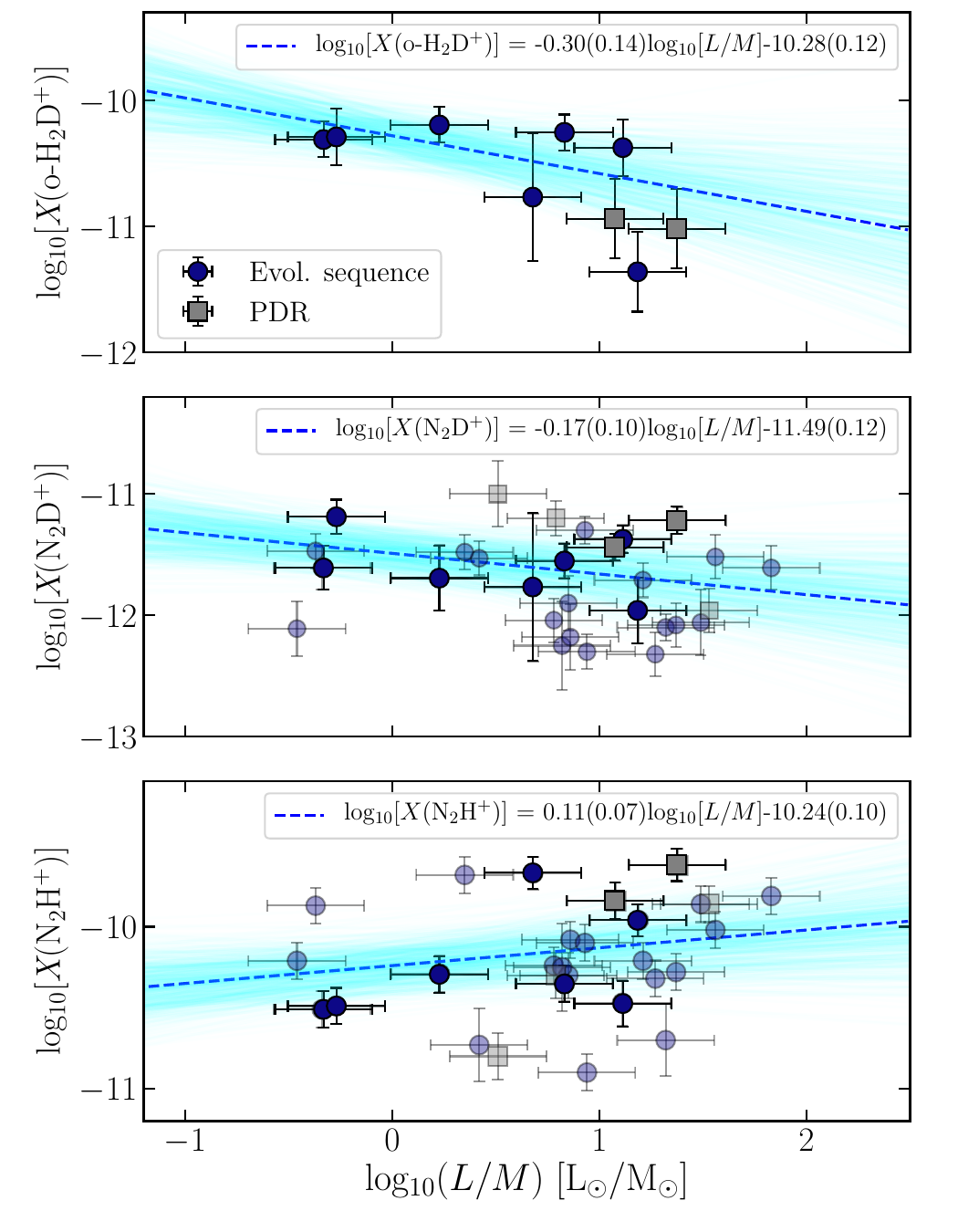}}
	\caption{Correlations between $L/M$ and the abundances of \ohhdp, \nndp, and N$_2$H$^+$ obtained with the LTE analysis. Blue dots refer to sources associated with the evolutionary sequence defined in Sect.~\ref{sec2.2:subsample}, whilst grey dots are associated with PDRs. Shaded dots refer to sources in which \ohhdp is not detected. The blue dashed lines show the result of a linear least-squares fit to the data, while the cyan shaded areas show the 3$\sigma$ uncertainties on the best-fit parameters (errors in parenthesis).}\label{fig:N_LM}
\end{figure}

\begin{figure}
	\centering
	{\includegraphics[width=1\columnwidth]{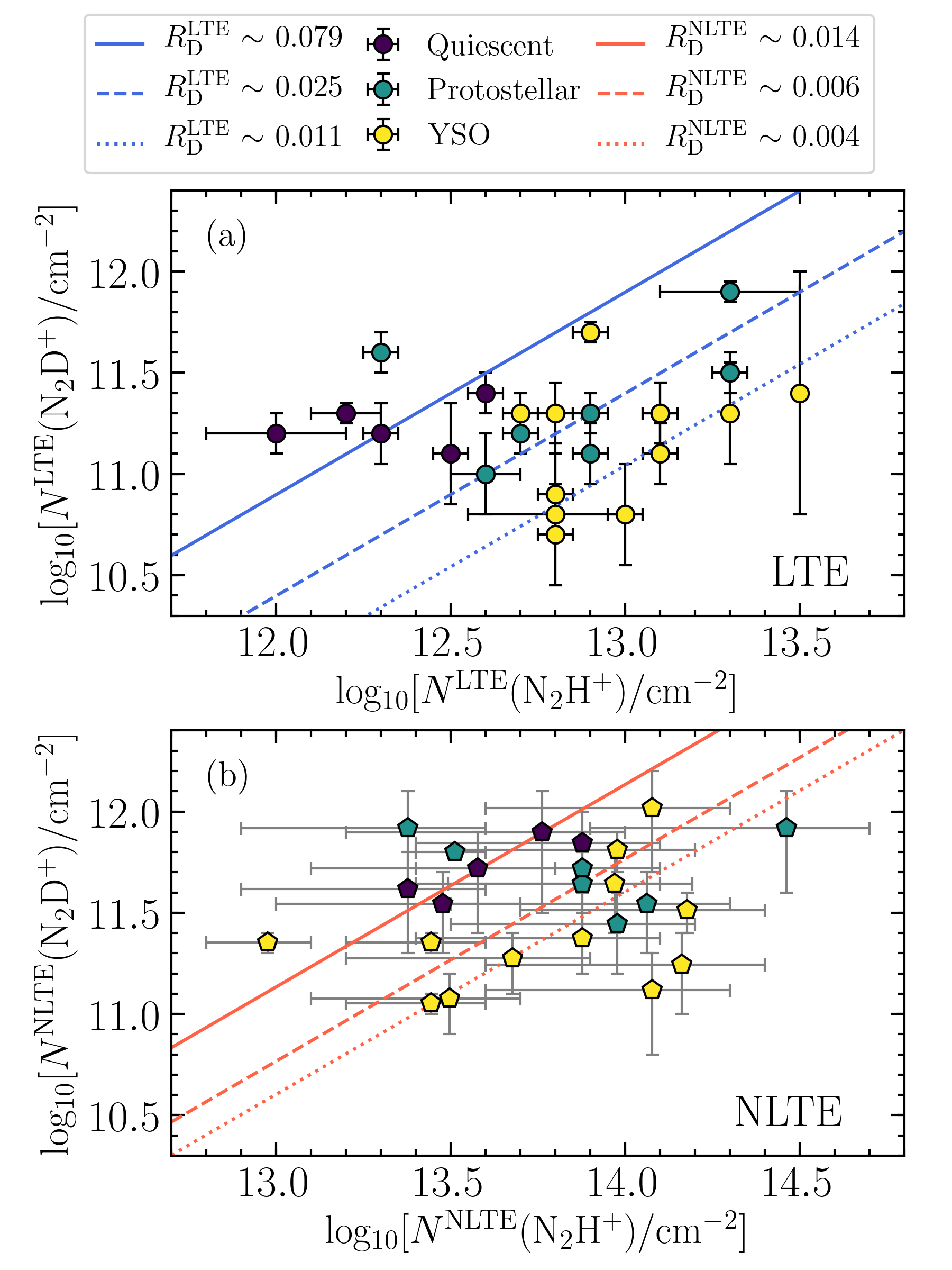}}
	\caption{Column densities of \nndp compared to those of N$_2$H$^+$ and obtained under (a) LTE and (b) NLTE conditions. The different colours of markers refer to the evolutionary stages of clumps, as in Fig.~\ref{fig:FWHM}. Lines show the median values of the deuteration fraction for the quiescent group (solid line; \mbox{$R^{\rm LTE}_{\rm D}$ = 0.079$\pm$0.029} and \mbox{$R^{\rm NLTE}_{\rm D}$ = 0.014$\pm$0.005}), the protostellar group (dashed line; \mbox{$R^{\rm LTE}_{\rm D}$ = 0.025$\pm$0.006} and \mbox{$R^{\rm NLTE}_{\rm D}$ = 0.006$\pm$0.001}), and the YSO group (dotted line; \mbox{$R^{\rm LTE}_{\rm D}$ = 0.011$\pm$0.004} and \mbox{$R^{\rm NLTE}_{\rm D}$ = 0.004$\pm$0.002}). The black bars show the uncertainties associated with $N^{\rm LTE}$, while the grey bars refer to the ranges in $N^{\rm NLTE}$ listed in Table~\ref{tab:summary}.}\label{fig:Rd}
\end{figure}

\subsubsection{Observed evolutionary trends}
Figure~\ref{fig:N_LM} shows how the LTE-abundances of o-H$_2$D$^+$, \nndp, and N$_2$H$^+$ vary across the different evolutionary stages, identified by the $L/M$ ratio. A linear least-squares fit to the data is shown in each panel as a blue dashed line. Remarkably, $X$(o-H$_2$D$^+$) decreases by $\sim$one orders of magnitude with increasing $L/M$, proving further evidence that \ohhdp is sensitive to the evolutionary stage of massive star formation. The abundance of N$_2$D$^+$ shows a similar downward trend over the same $L/M$ range, which is consistent with the results obtained by \cite{Fontani11} in similar environments. Finally, $X$(N$_2$H$^+$) gradually increases with increasing $L/M$, suggesting that the formation of N$_2$H$^+$ is favoured at later stages, possibly correlating with the higher densities and temperatures associated with increased $L/M$ (see Sect.~\ref{sec5:dicussion}). Finally, we note that the trends observed for $X$(o-H$_2$D$^+$) and $X$(N$_2$H$^+$) are also confirmed when considering the variation of $N$(o-H$_2$D$^+$) and $N$(N$_2$H$^+$) with $L/M$, while $N$(N$_2$D$^+$) remains flat with increasing $L/M$.\\
\indent As a consequence of the above trends, it was also found that the deuteration fraction derived from diazenylium considering the N$_2$D$^+$(3-2) and N$_2$H$^+$(4-3) rotational transitions, namely, \mbox{$R_{\rm D} = N({\rm N_2D^+})/N({\rm N_2H^+})$}, varies as a function of the different evolutionary classes. This is consistent with previous studies (e.g. \citealt{Chen11, Fontani11}), where $R_{\rm D}$ increases with decreasing temperature and increasing gas density. Figure~\ref{fig:Rd}a compares $N^{\rm LTE}$(N$_2$H$^+$) with $N^{\rm LTE}$(N$_2$D$^+$) across the different evolutionary stages. We note a significant difference in the median deuteration fraction for the different evolutionary stages of massive star-forming clumps. The quiescent stage shows the highest median $R^{\rm LTE}_{\rm D}$ (=0.079$\pm$0.029), the protostellar stage exhibits intermediate values (=0.025$\pm$0.006), while the YSO stage displays the lowest $R^{\rm LTE}_{\rm D}$ (=0.011$\pm$0.004). In the latter case, the $R_{\rm D}$ is, on average, an order of magnitude lower than for quiescent sources. To confirm the statistical significance of the observed separation between the evolutionary classes, we applied the Anderson-Darling (AD; e.g. \citealt{Scholz87}) test to the distribution of $R_{\rm D}$. We found a statistically significant difference in the $R_{\rm D}$ distribution between quiescent and protostellar stages and between the protostellar and YSO, with a low probability ($\mathcal{P}$$<$0.02) that the two samples originate from the same population. This separation is even more significant when considering the quiescent and YSO classes ($\mathcal{P}$$\sim$0.001), indicating a clear chemical diversity and evolution across these stages. A similar, albeit less pronounced trend is observed for the median $R^{\rm NLTE}_{\rm D}$ derived from the NLTE analysis (see Fig.~\ref{fig:Rd}b).

\section{Discussion}\label{sec5:dicussion}
\begin{figure}
	\centering
	{\includegraphics[width=1\columnwidth]{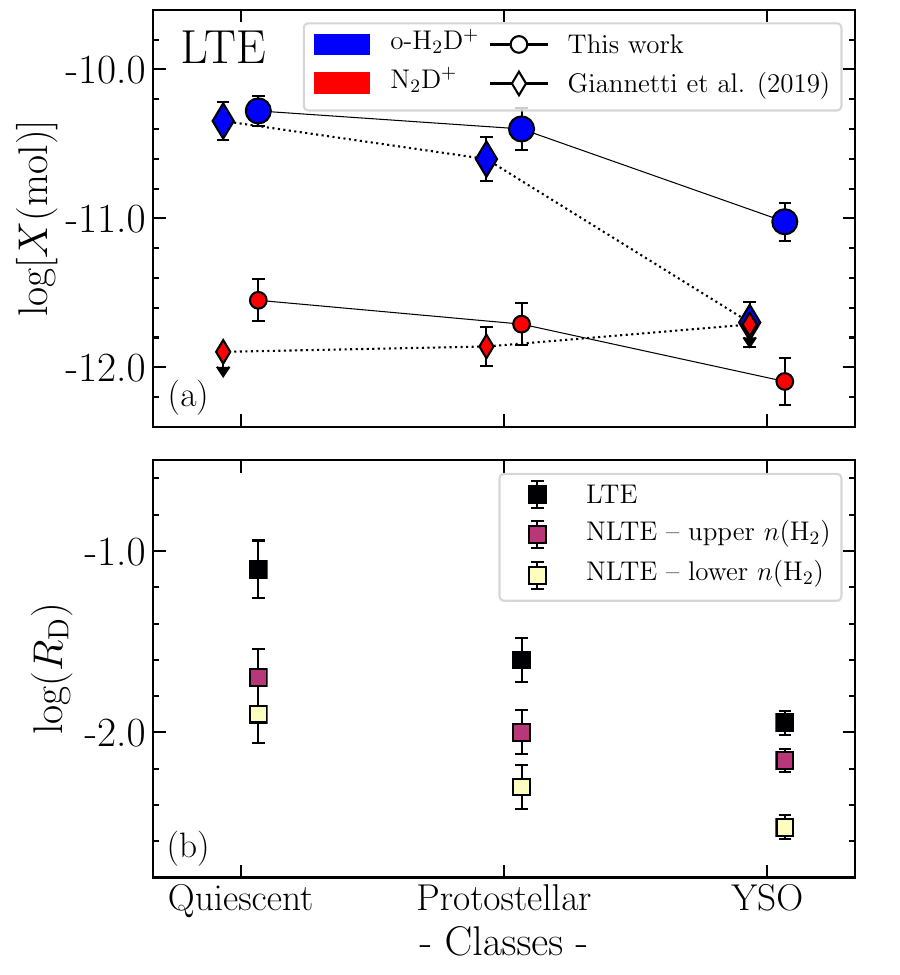}}
	\caption{Summary of the observed evolutionary trends. Panel (a): Abundances of o-H$_2$D$^+$ (blue symbols) and N$_2$D$^+$ (red symbols) as a function of the evolutionary class defined in Sect.~\ref{sec2.2:subsample}. Circles and diamonds refer to results from this work and \cite{Giannetti19}, respectively. Panel (b):  Median $R_{\rm D}$ factors derived for each evolutionary class defined in our sample. Different colours refer to values obtained from LTE and NLTE analysis discussed in Sects.~\ref{subsec:LTEcol} and~\ref{subsec:nonLTEcol}.}\label{fig:abundances}
\end{figure}

Figure~\ref{fig:abundances}a shows the variation of the average abundances of \mbox{\ohhdp}and N$_2$D$^+$ resulting from the LTE analysis, as a function of the evolutionary class (see Sect.~\ref{sec2.2:subsample}). When considering the o-H$_2$D$^+$ (blue markers), our results are in agreement with \cite{Sabatini20}, and confirm a decreasing trend in \Xo as clumps evolution progresses. 
However, the new classification and temperatures assumed to derive $N^{\rm LTE}$(o-H$_2$D$^+$) lead to a small variation in the median \Xo ratios among the different stages. The quiescent-to-protostellar ratio decreases from $\sim$2.0 in \cite{Sabatini20} to $\sim$1.3 in Fig.~\ref{fig:abundances}a. Conversely, the protostellar-to-YSO ratio increases from $\sim$4.0 to $\sim$4.3. The overall downward trend of \Xo with evolution remains unchanged, confirming that the new classification has only a minor impact on the final results. The ratio between the median values of \Xo in quiescent and YSO sources is also comparable to previous results (a factor of $\sim$6 in Fig.~\ref{fig:abundances}a compared to $\sim$8 in \citealt{Sabatini20}). This difference is consistent with the estimated uncertainties and with the results by \cite{Giannetti19}, i.e. blue diamonds in Fig.~\ref{fig:abundances}a.\\
\indent The red markers in Fig.~\ref{fig:abundances}a show the trend of \Xn as a function of evolutionary classes. In contrast to \cite{Giannetti19}, our findings reveal a slight decrease in \Xn with progressing star formation, with a difference of a factor of $\sim$2.9 between the quiescent and YSO stages. However, it is important to note several caveats regarding this comparison: (1) Although the physical properties of the clumps and the observational setup (e.g. telescope, spectral setup, sensitivity) in \cite{Giannetti19} are similar to those adopted here, the former sample is small (three clumps) and not statistically representative, as all three clumps belong to the same filamentary IRDC, G351.77-0.51 \citep[e.g.][]{Sabatini19}. (2) One of the three sources in \cite{Giannetti19} presents only an upper limit for $X$(N$_2$D$^+$), making it is difficult to assess evolutionary trends. The anticorrelation between the abundances of \ohhdp and \nndp -- and, consequently, the reliability in using the \Xo/\Xn ratio as an evolutionary indicator of high-mass star-forming clumps -- are not confirmed in our sample.\\
\indent To interpret the results presented in Fig.~\ref{fig:abundances}a, we consider both chemical evolution and beam dilution effects that may affect our observational results. During the early quiescent phase, \ohhdp is expected to be more abundant than N$_2$D$^+$, as the former is the initial product of deuterium fractionation \citep[e.g.][]{Dalgarno84, Caselli12b, Ceccarelli14}, while the formation of \nndp is limited by the presence of H$_2$D$^+$ and N$_2$ \citep[e.g.][]{Oberg21}. The formation of both H$_2$D$^+$ and N$_2$D$^+$  is significantly influenced by the abundance of CO in the gas phase, which is strongly depleted in our quiescent sample with an average CO depletion factor ($\fD$)\footnote{Defined as the ratio between the expected CO/H$_2$ abundance and the observed one (e.g., \citealt{Caselli99}).} of 6 \citep{Sabatini20}. As the evolution progresses (protostellar stage), the \ohhdp can transform into its multiply deuterated isotopologues (D$_2$H$^+$ and D$_3^+$), possibly contributing to the observed slight decrease in \Xo. At this stage, the average CO depletion factor remains similar to that of quiescent sources \citep[see][]{Giannetti14} and any effect of heating on the chemistry is still negligible on the clump scales, as it occurs on very small scales as shown by \cite{Sabatini22}. The decrease in \Xo would consequently also affect \Xn, as the chemistry of \nndp is strongly connected with that of H$_2$D$^+$. Furthermore, N$_2$ depletion could further reduce the N$_2$D$^+$ abundance \citep[e.g.][]{Redaelli19, Sipila19}. During the YSO phase, the temperature increase leads to CO desorption from the dust grains, resulting in the destruction of both o-H$_2$D$^+$ and N$_2$D$^+$ near the YSO. In addition, a higher ortho-to-para-H$_2$ ratio suppresses deuterium fractionation (e.g. \citealt{Hugo09}). These factors lead to a significant reduction of the \ohhdp abundance. Although \nndp may also be affected, its overall decrease is less pronounced, probably because its emission arises not only from the neighbourhood of the YSO, but also from the outer envelope, where \ohhdp continues to transform into heavier isotopologues, as modelled by \cite{Sipila16}. We find that with time, \Xn evolves similarly to \Xo but also that the former appears to be less sensitive to density and temperature variations that typically affect deuterated species during star formation, at least on the clump scale. The NLTE analysis discussed in Sect.~\ref{subsec:nonLTEcol} further supports this scenario and shows that, among the analysed tracers, the column density of \ohhdp is the least sensitive to the assumed gas density. This is in line with the results of \cite{Miettinen20}, who also found that it is difficult to detect a clear trend in $X$(N$_2$D$^+$), suggesting that this molecule is not ideal to follow the star formation process and may also lead to misleading results when used as a proxy for the identification of prestellar cores.\\ 
\indent The only other evolutionary indicator we find is the $\Rd$ determined from \nnhp and \nndp(see Fig.~\ref{fig:Rd}). Regardless of the method used to derive the column density of each tracers, we observe that $\Rd$ decreases by almost one order of magnitude between quiescent and YSO stages (see Fig.\ref{fig:abundances}b). This trend is similar to what was found for \Xo, and suggests that the physical conditions regulating the formation and destruction of deutereted molecules evolve consistently during the low- and high-mass star formation \citep[see][]{Chen11, Fontani11}. However, caution is also required when comparing the observed chemical properties in low- and high-mass star-forming regions, especially due to beam dilution effects. \cite{Sabatini22} show that validity of typical evolutionary indicators, such as $\fD$, can vary significantly depending on the spatial scale considered. Based on the pilot sample of the ALMA Survey of 70~$\mu$m Dark High-mass Clumps in Early Stages (ASHES; e.g. \citealt{Sanhueza19, Li23}), \cite{Sabatini22} found that while $\fD$ decreases significantly on a large scale with increasing $L/M$ of the clumps, the population of protostellar cores embedded in the ASHES clumps display $\fD$ that are, on average, higher than in prestellar cores. This is due to the fact that the CO-depletion process in the cold outer envelopes of the cores is still efficient even when a young stellar object has already formed. The same effect appears to be less pronounced in the low-mass star-forming regions, as the high-density envelopes are less extended, and more affected by the temperature fluctuations caused by the star formation activity. This affects also deuterium fractionation, which is only efficient in an environment strongly depleted of neutrals, and in particular CO. At the same time, the calculation of $\Rd$  assumes that both N$_2$H$^+$ and N$_2$D$^+$ are co-spatially distributed along the line of sight. However, interferometric observations reveal that N$_2$D$^+$ typically traces more compact regions than N$_2$H$^+$, especially in high-mass star formation, not always coinciding with the cores identified via \ohhdp \citep[e.g.][]{Redaelli22}. We find that this effect is more pronounced in evolved clumps, as shown by the consistently broader \nnhp line profiles compared to \nndp(Fig.~\ref{fig:FWHM}). Thus, while the mean $\Rd$ can serve as a reliable evolutionary indicator on clump scales, it may not accurately reflect the actual deuteration fraction of cores. This is especially important for the identification of prestellar cores in HMSFRs, as $\Rd$ is often used as an indicator for these cold, dense, prestellar environments.

\section{Conclusions}\label{sec6:conclusion}
To assess the validity of the \Xo/\Xn ratio as an evolutionary indicator for the high-mass star formation process, we conducted APEX observations of \nnhp(4-3) and \nndp(3-2) in a sample of 40 massive clumps at different evolutionary stages selected from the ATLASGAL survey \citep{Schuller09, Urquhart22}. The results for \nnhp and \nndp were compared with the \ohhdp\mbox{(1$_{10}$ – 1$_{11}$)} detections reported by \cite{Sabatini20}. We found emission of \nnhp in all observed clumps, while \nndp was detected in 29 out of 40 sources, with decreasing detection rates with clump evolution, similar to \mbox{o-H$_2$D$^+$}. Quantitatively, the detection rate of \nndp decreases from 100\% to 67\%, while that of \ohhdp ranges from 80\% to 10\%, from the quiescent to the YSO stages, respectively (Table~\ref{tab:observations}).\\
\indent As the clumps evolve, the \nnhp lines become broader, passing from typical FWHMs of $\sim$2.5~km~s$^{-1}$ in quiescent clumps to $\sim$3.6~km~s$^{-1}$ in YSO and exhibit multiple velocity components tracing larger, more turbulent structures in more evolved sources. Conversely, the \nndp lines are on average narrower than \nnhp lines, with FWHMs in the $\sim$1.8-2.3~km~s$^{-1}$ range, which suggest that they are associated with the denser and cooler cores into the clumps along the whole evolutionary sequence.\\
\indent The abundances of \ohhdp and \nndp decrease from (5.3$\pm$1.2)$\times$10$^{-11}$ to (1$\pm$0.3)$\times$10$^{-11}$ and from (2.8$\pm$0.9)$\times$10$^{-12}$ to (0.8$\pm$0.3)$\times$10$^{-12}$, respectively, following the clumps' evolution, which is consistent with previous findings \citep[e.g.,][see Fig.~\ref{fig:abundances}a]{Chen11, Fontani11, Sabatini20}. The validity of the \Xo/\Xn as an evolutionary indicator is not confirmed in our sample. However, \Xo shows a much stronger decrease (a factor of $\sim$6) over the different evolutionary stages than the \nndp abundance (a factor of $\sim$3). At the same time, when considering the column densities of the different tracers, this trend is confirmed only in the case of \mbox{o-H$_2$D$^+$}, while $N$(N$_2$D$^+$) shows flat values with increasing $L/M$ of clumps.\\ 
\indent The deuteration fraction resulting from \nnhp and \nndp is also confirmed as a potential evolutionary indicator in massive star-forming clumps, as it decreases by almost an order of magnitude between quiescent and YSO stages. However, due to beam dilution effects that affect the individual observation, the interpretation of the latter indicator is more complex as the emission of \nndp typically maps more compact regions than N$_2$H$^+$ \citep[e.g.][]{Miettinen11, Redaelli22}.\\
\indent In conclusion, our study confirms the potential of o-H$_2$D$^+$ and $\Rd$ as tracers of the different evolutionary stages in high-mass star formation process. High-resolution interferometric observations are now crucial to accurately calibrate these indicators by resolving the denser cores harboured into massive clumps to mitigate beam dilution effects.

\begin{acknowledgements}
	The authors thank the anonymous Referee for all the helpful suggestions that improved the manuscript, and Dr L. Podio for fruitful scientific discussions and feedback. This publication is based on data acquired with the Atacama Pathfinder Experiment (APEX) under programme IDs [0101.F-9517, M-097.F-0039-2016, M-098.F-0013-2016, C-0105.F-9715C-2020 and C-0107.F-9711-2021]. APEX has been a collaboration between the Max-Planck-Institut fur Radioastronomie, the European Southern Observatory, and the Onsala Space Observatory. GS acknowledges the project PRIN-MUR 2020 MUR BEYOND-2p (``Astrochemistry beyond the second period elements'', Prot. 2020AFB3FX), and the INAF-Minigrant 2023 TRIESTE (``TRacing the chemIcal hEritage of our originS: from proTostars to planEts''; PI: G. Sabatini). SB acknowledges BASAL Centro de Astrofisica y Tecnologias Afines (CATA), project number AFB-17002. This research has made use of the IRAM GILDAS software (\url{http://www.iram.fr/IRAMFR/GILDAS}), the Cologne Database for Molecular Spectroscopy (CDMS), the NASA’s Astrophysics Data System Bibliographic Services (ADS), Astropy (\citealt{Astropy13, Astropy18}; see also \url{http://www.astropy.org}) and Matplotlib (\citealt{Matplotlib07}).
\end{acknowledgements}

%
%
\bibliographystyle{aa} 
\bibliography{mybib_GAL}

\begin{appendix}
	
\begin{table*}	
	\section{Physical and chemical properties of the clumps}
	\label{sec:appB}
	\indent Table~\ref{tab:comparison} presents a comparison between the main physical properties and evolutionary classifications of clumps with \ohhdp detection. The left columns display values from \cite{Konig17} as adopted by \cite{Sabatini20}, while the right columns show the recently updated values from \cite{Urquhart22}. For ease of comparison, we have assigned numerical identifiers to the different evolutionary stages. Uncertainties associated with the  $N^{\rm LTE}$(o-H$_2$D$^+$) estimates are derived from the fiducial fit provided by {\verb~MCWeeds~}.\\
	\indent Table~\ref{tab:summary} summarises the physical properties of all the clumps considered in this work, and the spectral properties and column densities of o-H$_2$D$^+$, \nnhp and \nndp derived with {\verb~MCWeeds~} ($N^{\rm LTE}$) and RADEX ($N^{\rm NLTE}$). Finally, the observed spectra of \nndp and N$_2$H$^+$ are shown in Fig.~\ref{fig:spectra} and discussed in Sect.~\ref{sec4:results}, while we refer to \cite{Sabatini20} for a comprehensive description of the \ohhdp APEX data.

		\footnotesize
		\caption{\label{tab:comparison}Comparison between the physical properties presented in \cite{Sabatini20} and those in \cite{Urquhart22}.}
		\setlength{\tabcolsep}{2.5pt}
		\renewcommand{\arraystretch}{1.1}
		\centering
		\begin{tabular}{l|ccccc|ccccc}
			\hline
			Employed in  & \multicolumn{5}{c|}{\cite{Sabatini20}} & \multicolumn{5}{c}{\cite{Urquhart22} and this work}\\
			\hline
			ATLASGAL-ID& $T_{\rm dust}$\tablefootmark{a} &$(L/M)$\tablefootmark{a}&$N$(H$_2$)\tablefootmark{a}&$N^{\rm NLTE}$(o-H$_2$D$^+$)\tablefootmark{b}&Class\tablefootmark{a}& $T_{\rm dust}$\tablefootmark{b} &$(L/M)$\tablefootmark{b}&$N$(H$_2$)\tablefootmark{b}&$N^{\rm NLTE}$(o-H$_2$D$^+$)&Class\tablefootmark{b,c} \\
			&(K)& $L_\odot/M_\odot$ &\multicolumn{2}{c}{----~~log(cm$^{-2}$)~~----}& (IDs) &(K)& $L_\odot/M_\odot$ &\multicolumn{2}{c}{----~~log(cm$^{-2}$)~~----}& (IDs)\\
			\hline                                                                          
			G12.50-0.22   & 13.0$\pm$0.2 & 1.3  & 22.8 & 12.8$^{+0.1}_{-0.1}$ & 70w (1)& 13.0$\pm$0.2 & 0.5 & 22.8 & 12.5$^{+0.1}_{-0.1}$ & Quiescent (1)\\          
			G13.18+0.06   & 24.2$\pm$0.8 & 22.5 & 22.9 & 12.2$^{+0.3}_{-0.3}$ & 70w (1)& 20.3$\pm$0.9 & 23.6 & 23.0 & 12.0$^{+0.3}_{-0.3}$ & PDR   (5)\\
			G14.11--0.57  & 22.4$\pm$0.8 & 9.1  & 22.9 & 12.5$^{+0.3}_{-0.3}$ & IRw (2)& 15.8$\pm$0.5 & 4.8 & 23.2 & 12.4$^{+0.4}_{-0.4}$ & YSO   (3)\\
			G14.49--0.14  & 12.4$\pm$0.4 & 0.4  & 23.1 & 13.0$^{+0.1}_{-0.1}$ & 70w (1)& 16.6$\pm$2.6 & 6.8 & 23.0 & 12.7$^{+0.1}_{-0.1}$ & Quiescent (1)\\
			G14.63--0.58  & 22.5$\pm$0.4 & 11.1 & 23.0 & 12.4$^{+0.2}_{-0.2}$ & IRw (2)& 19.1$\pm$4.3 &11.9 & 23.1 & 12.2$^{+0.3}_{-0.3}$ & PDR   (5)\\
			G15.72--0.59  & 12.1$\pm$0.5 & 0.2  & 22.8 & 12.8$^{+0.2}_{-0.2}$ & IRw (2)& 12.1$\pm$0.5 & 0.6 & 22.9 & 12.5$^{+0.2}_{-0.2}$ & Protostellar (2)\\   
			G19.88--0.54  & 24.2$\pm$1.4 & 15.5 & 23.1 & 12.1$^{+0.3}_{-0.3}$ & IRb (3)& 20.3$\pm$2.8 &15.3 & 23.3 & 11.9$^{+0.2}_{-0.2}$ & YSO   (3)\\
			G351.57+0.76  & 17.0$\pm$0.1 & 2.7  & 22.7 & 12.6$^{+0.2}_{-0.2}$ & 70w (1)& 20.1$\pm$9.5 & 13.0 & 22.7 & 12.3$^{+0.2}_{-0.2}$ & Quiescent (1)\\
			G354.95--0.54 & 19.1$\pm$1.3 & 3.2  & 22.6 & 12.7$^{+0.1}_{-0.1}$ & 70w (1)& 15.0$\pm$1.7 & 1.7 & 22.8 & 12.6$^{+0.1}_{-0.1}$ & Quiescent (1)\\
			\hline
		\end{tabular}
		\tablefoot{\tablefoottext{a}{Data from \cite{Sabatini20}. Due to the selection criteria applied in this study on the TOP100 sources (see Sect.~\ref{sec2.2:subsample}), seven sources detected in \ohhdp were excluded from this analysis, given they are associated with a heliocentric distance $\geq$~4~kpc}; \tablefoottext{b}{Data from \cite{Urquhart22};} \tablefoottext{c}{Evolutionary IDs follow the classification in Sect.~\ref{sec2.2:subsample}.}}
	\end{table*}
	
	\begin{landscape}
		\begin{table}[p] 
			\scriptsize
			\caption{\label{tab:summary}Physical and spectral properties of sample presented in Sect.~\ref{sec2.2:subsample}.}
			\setlength{\tabcolsep}{1pt}
			\renewcommand{\arraystretch}{1.2}
			\centering
			\begin{tabular}{l|cccccccc|ccccc|ccccc|ccccc|cc}
				\hline\hline
				&                 &             &               &               &     &               &            &            & \multicolumn{5}{c|}{-----------------~~~~~ \ohhdp ~~~~~-----------------} & \multicolumn{5}{c|}{-----------------~~~~~ \nnhp ~~~~~------------------}& \multicolumn{5}{c|}{-----------------~~~~~ \nndp ~~~~~-----------------}  & \\
				ATLASGAL-ID & $D_{\rm \odot}$ &$D_{\rm GC}$ & $V_{\rm 0}$ & $T_{\rm dust}$ & $L/M$ & $R_{\rm eff}$ & $N$(H$_2$) & $n$(H$_2$)\tablefootmark{{\rm ({\it a})}} & $V_{\rm lsr}$ & FWHM & $N^{\rm LTE}$ & $N^{\rm NLTE}$ & $\tau^{\rm NLTE}$ & $V_{\rm lsr}$ & FWHM & $N^{\rm LTE}$ & $N^{\rm NLTE}$ & $\tau^{\rm NLTE}$ & $V_{\rm lsr}$ & FWHM & $N^{\rm LTE}$ & $N^{\rm NLTE}$ & $\tau^{\rm NLTE}$ & Class\\
				&\multicolumn{2}{c}{~~(kpc)~~}& (km~s$^{-1}$) & (K)& $L_\odot/M_\odot$ & (pc) & log(cm$^{-2}$) & log(cm$^{-3})$ & \multicolumn{2}{c}{~~(km~s$^{-1}$)~~} & \multicolumn{2}{c}{~~log(cm$^{-2}$)~~} & & \multicolumn{2}{c}{~~(km~s$^{-1}$)~~} & \multicolumn{2}{c}{~~log(cm$^{-2}$)~~} & & \multicolumn{2}{c}{~~(km~s$^{-1}$)~~} & \multicolumn{2}{c}{~~log(cm$^{-2}$)~~} & &\\
				\hline            
				G12.50-0.22                                 &2.6&5.8& 35.3 &13.0$\pm$0.2 &  0.5&0.20&22.81&[5.02, 5.72]& 36.0$^{+0.8}_{-0.5}$ & 1.2$^{+0.3}_{-0.3}$ & 12.5$^{+0.1}_{-0.1}$ &[12.7, 12.6]&[0.05, 0.03]&36.1$^{+0.1}_{-0.1}$ &1.9$^{+0.2}_{-0.2}$ &12.3$^{+0.1}_{-0.1}$&[13.8, 13.1]&[1.4, 0.4] & 35.6$^{+0.2}_{-0.2}$ & 1.1$^{+0.5}_{-0.4}$ & 11.2$^{+0.1}_{-0.2}$ &[11.9, 11.4] & [0.10, 0.04]& (1)\\
				G14.49-0.14                                 &3.1&5.4& 40.5 &16.6$\pm$2.8 &  6.8&0.26&22.95&[5.04, 5.74]& 39.8$^{+0.5}_{-0.1}$ & 2.8$^{+0.9}_{-0.8}$ & 12.7$^{+0.1}_{-0.1}$ &[12.9, 12.7]&[0.05, 0.03]&39.7$^{+0.1}_{-0.1}$ &3.3$^{+0.2}_{-0.1}$ &12.6$^{+0.1}_{-0.1}$&[14.1, 13.4]&[2.4, 0.5] & 39.5$^{+0.3}_{-0.3}$ & 2.8$^{+0.7}_{-0.7}$ & 11.4$^{+0.1}_{-0.1}$ &[12.0, 11.6] & [0.07, 0.02]& (1)\\
				G305.80-0.10                                &4.0&6.8&-41.2 &16.2$\pm$0.5 &  2.7&0.32&22.73&[4.74, 5.44]&  --                  &  --                 & --                   &--          &--          &-41.0$^{+0.6}_{-0.6}$ &3.4$^{+1.2}_{-1.0}$ &12.0$^{+0.2}_{-0.2}$&[14.0, 13.2]&[0.7, 0.2] &-41.3$^{+0.2}_{-0.3}$ & 2.0$^{+0.6}_{-0.4}$ & 11.2$^{+0.1}_{-0.1}$ &[12.1, 11.5] & [0.08, 0.03]& (1)\\
				G351.57+0.77                                &1.3&7.1&-2.7  &20.1$\pm$9.5 & 13.0&0.17&22.67&[4.95, 5.65]& -3.1$^{+0.7}_{-0.5}$ & 1.0$^{+0.5}_{-0.4}$ & 12.3$^{+0.2}_{-0.2}$ &[12.5, 12.3]&[0.03, 0.01]&-3.0$^{+0.1}_{-0.1}$ &1.6$^{+0.2}_{-0.2}$ &12.2$^{+0.1}_{-0.1}$&[13.6, 12.9]&[1.4, 0.4] &-3.0$^{+0.1}_{-0.1}$  & 0.7$^{+0.2}_{-0.1}$ & 11.3$^{+0.1}_{-0.1}$ &[11.8, 11.3] & [0.15, 0.05]& (1)\\
				G354.95-0.54                                &1.4&7.0&-5.5  &15.0$\pm$1.7 &  1.7&0.11&22.79&[5.26, 5.96]& -5.9$^{+0.1}_{-0.1}$ & 1.2$^{+0.4}_{-0.3}$ & 12.6$^{+0.1}_{-0.1}$ &[12.7, 12.6]&[0.06, 0.04]&-5.8$^{+0.2}_{-0.2}$ &2.4$^{+0.1}_{-0.1}$ &12.5$^{+0.1}_{-0.1}$&[13.7, 13.0]&[1.2, 0.3] &-5.9$^{+0.5}_{-0.5}$  & 1.9$^{+1.3}_{-1.0}$ & 11.1$^{+0.3}_{-0.2}$ &[11.7, 11.3] & [0.05, 0.02]& (1)\\
				\hline                                  
				G14.19-0.19                                 &3.1&5.4&38.7  &16.1$\pm$1.8 &  6.0&0.23&23.14&[5.29, 5.99]&  --  &  -- & --                   &--          &--          &38.9$^{+0.2}_{-0.2}$ &4.3$^{+0.3}_{-0.1}$ &12.9$^{+0.1}_{-0.1}$&[14.2, 13.5]&[2.4, 0.6] & 39.1$^{+0.2}_{-0.2}$ & 1.6$^{+0.5}_{-0.5}$ & 11.1$^{+0.2}_{-0.1}$ &[11.6, 11.2] & [0.04, 0.02]& (2)\\
				G15.72-0.59                                 &1.8&6.6& 17.8 &12.1$\pm$0.5 &  0.6&0.12&22.90&[5.22, 5.92]& 17.7$^{+0.9}_{-0.7}$ & 2.0$^{+0.2}_{-0.1}$ & 12.5$^{+0.1}_{-0.1}$ &[12.7, 12.6]&[0.04, 0.03]&17.6$^{+0.1}_{-0.1}$ &1.8$^{+0.2}_{-0.3}$ &12.3$^{+0.1}_{-0.1}$&[13.6, 12.9]&[0.8, 0.3] & 17.5$^{+0.1}_{-0.1}$ & 1.4$^{+0.3}_{-0.2}$ & 11.6$^{+0.1}_{-0.1}$ &[12.1, 11.6] & [0.15, 0.06]& (2)\\
				G22.37+0.45                                 &3.6&5.2&52.5  &13.1$\pm$0.2 &  0.6&0.18&22.95&[5.21, 5.91]&  --  &  -- & --                   &--          &--          &52.6$^{+0.1}_{-0.3}$ &3.6$^{+0.3}_{-0.2}$ &13.0$^{+0.1}_{-0.1}$&[14.5, 13.7]&[6.2, 1.1] & --   & --  & --                   &          -- & --          & (2)\\
				G305.19-0.01                                &4.0&6.9&-34.0 &23.1$\pm$5.5 & 16.1&0.28&22.91&[4.97, 5.67]&  --  &  -- & --                   &--          &--          &-34.4$^{+0.2}_{-0.2}$&4.2$^{+0.3}_{-0.3}$ &12.7$^{+0.1}_{-0.1}$&[14.1, 13.4]&[2.2, 0.5] & -34.0$^{+0.5}_{-0.5}$ & 3.5$^{+0.5}_{-0.8}$ & 11.2$^{+0.1}_{-0.1}$ &[11.8, 11.4] & [0.03, 0.01]& (2)\\
				G326.99-0.03                                &3.5&5.8&-59.4 &16.7$\pm$0.5 &  2.3&0.18&22.98&[5.24, 5.94]&  --  &  -- & --                   &--          &--          &-58.6$^{+0.1}_{-0.1}$&4.2$^{+0.2}_{-0.3}$&13.3$^{+0.1}_{-0.1}$&[14.7, 13.9]&[11.2, 2.0]& -58.2$^{+0.3}_{-0.3}$ & 2.5$^{+0.8}_{-0.7}$ & 11.5$^{+0.1}_{-0.1}$ &[12.1, 11.6] & [0.08, 0.03]& (2)\\
				G340.37-0.39                                &3.6&5.1&-44.4 &16.1$\pm$0.2 &  7.1&0.25&22.90&[5.01, 5.71]&  --  &  -- & --                   &--          &--          &-44.0$^{+0.2}_{-0.2}$&3.7$^{+0.5}_{-0.4}$ &12.6$^{+0.1}_{-0.1}$&[14.1, 13.4]&[1.9, 0.4] & -43.9$^{+0.7}_{-0.8}$& 2.9$^{+1.1}_{-1.1}$ & 11.0$^{+0.2}_{-0.2}$ &[11.9, 11.4] & [0.04, 0.02]& (2)\\
				\multirow{2}{*}{G351.45+0.66$^{\dagger}$}   &1.3&7.1&-3.9  &21.4$\pm$1.7 & 20.8&0.12&23.97&[6.40, 7.10]&  --  &  -- & --                   &--          &--          &-1.5$^{+0.1}_{-0.1}$& 3.3$^{+0.1}_{-0.1}$ & (c1) 13.3$^{+0.1}_{-0.1}$& [13.6, 13.4] & [1.2, 0.7] & -4.3$^{+0.1}_{-0.1}$ & 3.2$^{+0.3}_{-0.2}$ & 11.9$^{+0.1}_{-0.1}$ &[11.8, 11.8] & [0.02, 0.02]& \multirow{2}{*}{(2)}\\
				& & & & & & & & & & & & & & -5.7$^{+0.1}_{-0.1}$ & 3.2$^{+0.1}_{-0.1}$ & (c2) 13.1$^{+0.1}_{-0.1}$& [13.7, 13.5] & [1.8, 0.9] &  & & & & &\\
				G353.07+0.45                                &1.4&7.0& 0.7  &16.3$\pm$0.6 &  0.4&0.12&22.77&[5.20, 5.90]&  --  &  -- & --                   &--          &--          &1.2$^{+0.2}_{-0.2}$  &2.6$^{+0.2}_{-0.2}$ &12.9$^{+0.1}_{-0.1}$&[14.3, 13.5]&[6.3, 1.1] & 1.3$^{+0.2}_{-0.1}$ & 1.2$^{+0.4}_{-0.3}$ & 11.3$^{+0.1}_{-0.1}$ &[11.7, 11.3] & [0.08, 0.03]& (2)\\
				\hline
				G14.11-0.57                                 &1.5&6.9& 20.1 &15.8$\pm$0.5 &  4.8&0.11&23.17&[5.64, 6.34]& 19.9$^{+0.1}_{-0.3}$ & 1.7$^{+0.3}_{-0.5}$ & 12.4$^{+0.4}_{-0.4}$ &[12.5, 12.4]  & [0.03, 0.02] &20.0$^{+0.1}_{-0.1}$ &3.0$^{+0.1}_{-0.1}$ &13.5$^{+0.1}_{-0.1}$&[14.2, 13.5]&[4.9, 1.1] & 20.2$^{+0.6}_{-0.7}$ & 3.6$^{+0.4}_{-0.8}$ & 11.4$^{+0.6}_{-0.6}$ &[11.9, 11.7] & [0.05, 0.02]& (3)\\
				G19.88-0.54                                 &3.3&5.3& 43.2 &20.3$\pm$2.8 & 15.3&0.16&23.26&[5.57, 6.27]& 43.7$^{+1.1}_{-0.9}$ & 1.3$^{+0.4}_{-0.2}$ & 11.9$^{+0.2}_{-0.2}$ &[12.0, 11.9]&[0.02, 0.01]&43.8$^{+0.1}_{-0.1}$ &3.6$^{+0.1}_{-0.1}$ &13.3$^{+0.1}_{-0.1}$&[14.4, 13.7]&[7.8, 1.6] & 43.6$^{+1.0}_{-1.0}$ & 3.1$^{+1.0}_{-1.0}$ & 11.3$^{+0.3}_{-0.2}$ &[11.6, 11.4] & [0.03, 0.01]& (3)\\
				G34.41+0.23                                 &2.9&6.2& 57.9 &22.7$\pm$4.4 & 23.1&0.12&23.38&[5.81, 6.51]&  --  & --  & --                   &--          &--          &57.6$^{+0.1}_{-0.1}$ &4.3$^{+0.1}_{-0.1}$ &13.1$^{+0.1}_{-0.1}$&[13.6, 13.2]&[1.0, 0.4] & 57.2$^{+0.2}_{-0.2}$ & 1.7$^{+0.5}_{-0.4}$ & 11.3$^{+0.2}_{-0.1}$ &[11.4, 11.3] & [0.03, 0.01]& (3)\\
				G34.82+0.35                                 &3.4&5.9& 56.9 &21.6$\pm$0.8 & 27.6&0.27&22.79&[4.87, 5.57]&  --  & --  & --                   &--          &--          &56.4$^{+0.1}_{-0.1}$ &3.9$^{+0.2}_{-0.2}$ &12.7$^{+0.1}_{-0.1}$&[14.4, 13.7]&[6.2, 1.1] & --   & --  & --                   &          -- & --          & (3)\\
				G35.20-0.74\tablefootmark{{\rm ({\it b})}}  &2.2&6.7& 33.9 &29.5$\pm$2.7 & 50.9&0.13&23.21&[5.61, 6.31]&  --  & --  & --                   &--          &--          &33.7$^{+0.1}_{-0.1}$ &4.9$^{+0.1}_{-0.1}$ &12.8$^{+0.1}_{-0.1}$&[13.5, 13.0]&[0.6, 0.2] & --   & --  & --                   &          -- & --          & (3)\\
				G53.14+0.07                                 &4.0&6.8& 21.8 &21.8$\pm$2.6 &  0.3&0.17&23.01&[5.29, 5.99]&  --  & --  & --                   &--          &--          &21.6$^{+0.1}_{-0.1}$ &2.7$^{+0.1}_{-0.1}$ &12.8$^{+0.1}_{-0.1}$&[13.9, 13.2]&[2.9, 0.7] & 21.7$^{+0.5}_{-0.4}$ & 1.7$^{+1.7}_{-1.0}$ & 10.9$^{+0.2}_{-0.2}$ &[11.4, 11.1] & [0.03, 0.01]& (3)\\
				G310.01+0.39                                &3.2&6.7&-41.7 &28.0$\pm$11.6&120.0&0.18&22.80&[5.06, 5.76]&  --  & --  & --                   &--          &--          &-41.2$^{+0.5}_{-0.5}$&2.4$^{+1.0}_{-0.8}$ &12.0$^{+0.1}_{-0.2}$&[13.3, 12.6]&[0.3, 0.1] & --   & --  & --                   &          -- & --          & (3)\\
				\multirow{2}{*}{G317.87-0.15$^{\dagger}$}   &2.3&6.8&-40.3 &19.4$\pm$1.2 &  6.5&0.10&23.05&[5.56, 6.26]&  --  & --  & --                   &--          &--          & -39.1$^{+0.1}_{-0.1}$ & 2.3$^{+0.3}_{-0.3}$ & (c1) 12.3$^{+0.2}_{-0.1}$& [13.7, 13.1] &[2.0, 0.6] & -40.0$^{+1.0}_{-0.9}$ & 2.1$^{+1.4}_{-1.4}$ & 10.8$^{+0.3}_{-0.4}$ &[11.2, 10.9] & [0.01, 0.01]& \multirow{2}{*}{(3)}\\
				&  & & & & & & & & & & & & & -41.4$^{+0.3}_{-0.3}$ & 3.8$^{+0.1}_{-0.3}$ & (c2) 12.7$^{+0.1}_{-0.1}$ & [13.9, 13.3] & [1.9, 0.5] & &  & & & & &\\
				G318.78-0.14                                &2.3&6.8&-37.7 &18.2$\pm$2.2 &  7.2&0.21&22.88&[5.07, 5.77]&  --  & --  & --                   &--          &--          &-38.4$^{+0.1}_{-0.2}$&4.0$^{+0.3}_{-0.3}$ &12.8$^{+0.1}_{-0.1}$&[14.3, 13.6]&[3.8, 0.7] & -37.4$^{+0.4}_{-0.7}$ & 1.4$^{+1.3}_{-0.9}$ & 10.7$^{+0.2}_{-0.3}$ &[11.3, 10.8] & [0.02, 0.01]& (3)\\
				G326.66+0.52                                &1.8&6.9&-39.6 &22.7$\pm$1.4 & 67.5&0.11&22.91&[5.38, 6.08]&  --  & --  & --                   &--          &--          &-39.4$^{+0.1}_{-0.1}$&3.3$^{+0.1}_{-0.1}$&13.1$^{+0.1}_{-0.1}$&[14.1, 13.4]&[4.3, 0.9] & -40.0$^{+0.3}_{-0.3}$& 2.0$^{+0.8}_{-0.7}$ & 11.3$^{+0.1}_{-0.2}$ &[11.5, 11.2] & [0.03, 0.01]& (3)\\
				G333.31+0.11                                &3.6&5.4&-47.0 &23.3$\pm$1.9 & 30.9&0.19&22.86&[5.09, 5.79]&  --  & --  & --                   &--          &--          &-46.7$^{+0.1}_{-0.1}$&3.8$^{+0.1}_{-0.1}$ &13.0$^{+0.1}_{-0.1}$&[14.4, 13.6]&[6.3, 1.1] & -46.8$^{+0.7}_{-0.7}$& 2.3$^{+1.4}_{-1.3}$ & 10.8$^{+0.2}_{-0.3}$ &[11.4, 11.0] & [0.02, 0.01]& (3)\\
				G339.62-0.12                                &2.6&6.0&-33.9 &24.6$\pm$4.2 & 38.2&0.18&22.80&[5.06, 5.76]&  --  & --  & --                   &--          &--          &-34.5$^{+0.1}_{-0.1}$&2.7$^{+0.1}_{-0.2}$ &12.7$^{+0.1}_{-0.1}$&[14.1, 13.3]&[3.8, 0.8] & --   & --  & --                   &          -- & --          & (3)\\
				G340.75-1.00                                &2.4&6.1&-29.1 &22.0$\pm$0.3 & 36.3&0.16&22.82&[5.13, 5.83]&  --  & --  & --                   &--          &--          &-29.2$^{+0.1}_{-0.1}$&3.4$^{+0.2}_{-0.2}$ &12.8$^{+0.1}_{-0.1}$&[14.1, 13.4]&[3.0, 0.7] & -29.3$^{+0.4}_{-0.5}$ & 2.7$^{+0.8}_{-0.9}$ & 11.3$^{+0.1}_{-0.2}$ &[11.8, 11.4] & [0.05, 0.02]& (3)\\
				G341.22-0.21                                &3.3&5.4&-44.3 &25.9$\pm$4.2 & 41.5&0.17&22.93&[5.21, 5.91]&  --  & --  & --                   &--          &--          &-43.8$^{+0.1}_{-0.1}$&3.7$^{+0.1}_{-0.1}$ &13.1$^{+0.1}_{-0.1}$&[14.3, 13.6]&[6.1, 1.1] & --   & --  & --                   &          -- & --          & (3)\\
				G343.75-0.16                                &2.6&5.9&-29.2 &20.8$\pm$1.2 & 18.5&0.09&23.42&[5.97, 6.67]&  --  & --  & --                   &--          &--          &-28.4$^{+0.1}_{-0.1}$&4.0$^{+0.1}_{-0.1}$ &13.1$^{+0.1}_{-0.1}$&[13.6, 13.2]&[0.9, 0.4] & -28.3$^{+0.3}_{-0.2}$ & 1.7$^{+0.6}_{-0.7}$ & 11.1$^{+0.1}_{-0.2}$ &[11.1, 11.0] & [0.01, 0.01]& (3)\\
				G344.23-0.57\tablefootmark{{\rm ({\it b})}} &2.2&6.2&-22.0 &22.0$\pm$1.3 &  8.7&0.11&23.57&[6.04, 6.74]&  --  & --  & --                   &--          &--          &-22.3$^{+0.2}_{-0.2}$&5.0$^{+0.5}_{-0.3}$ &12.7$^{+0.1}_{-0.1}$&[13.1, 12.8]&[0.2, 0.1] & -22.1$^{+0.4}_{-0.2}$ & 2.4$^{+0.4}_{-0.5}$ & 11.3$^{+0.1}_{-0.1}$ &[11.4, 11.3] & [0.01, 0.01]& (3)\\
				G345.51+0.35\tablefootmark{{\rm ({\it b})}} &2.4&6.1&-17.5 &32.7$\pm$1.5 &102.1&0.14&23.10&[5.47, 6.17]&  --  & --  & --                   &--          &--          &-17.0$^{+0.1}_{-0.1}$&4.1$^{+0.1}_{-0.1}$ &13.0$^{+0.1}_{-0.1}$&[13.8, 13.2]&[1.6, 0.4] & --   & --  & --                   &          -- & --          & (3)\\
				G345.72+0.82                                &2.4&6.1&-11.0 &18.6$\pm$6.1 &  8.4&0.21&23.00&[5.19, 5.89]&  --  & --  & --                   &--          &--          &-11.5$^{+0.1}_{-0.1}$&3.2$^{+0.1}_{-0.2}$ &12.9$^{+0.1}_{-0.1}$&[14.3, 13.6]&[6.2, 1.1] & -11.8$^{+0.1}_{-0.1}$ & 2.6$^{+0.3}_{-0.4}$ & 11.7$^{+0.1}_{-0.1}$ &[12.2, 11.7] & [0.10, 0.04]& (3)\\
				\hline
				G13.18+0.06                                 &2.6&5.8& 49.0 &20.3$\pm$0.9 & 23.6&0.20&23.02&[5.22, 5.92]& 48.5$^{+0.5}_{-0.4}$ & 1.3$^{+1.0}_{-0.3}$ & 12.0$^{+0.3}_{-0.3}$ &[12.2, 12.1]&[0.02, 0.01]&49.5$^{+0.1}_{-0.1}$ &4.8$^{+0.1}_{-0.1}$ &13.4$^{+0.1}_{-0.1}$&[14.5, 13.8]&[7.5, 1.4] & 49.1$^{+0.1}_{-0.1}$ & 2.0$^{+0.3}_{-0.2}$ & 11.8$^{+0.1}_{-0.1}$ &[12.1, 11.8] & [0.13, 0.04]& (4)\\
				G14.63-0.58                                 &1.5&6.9&18.2  &19.1$\pm$4.3 & 11.9&0.12&23.14&[5.57, 6.27]& 18.5$^{+0.5}_{-0.6}$ & 1.6$^{+0.9}_{-0.5}$ & 12.2$^{+0.3}_{-0.3}$ &[12.4, 12.2]&[0.02, 0.01]&18.3$^{+0.1}_{-0.1}$ &3.6$^{+0.1}_{-0.1}$ &13.3$^{+0.1}_{-0.1}$&[14.2, 13.5]&[4.3, 1.0] & 18.5$^{+0.1}_{-0.1}$ & 1.8$^{+0.2}_{-0.2}$ & 11.7$^{+0.1}_{-0.1}$ &[11.8, 11.6] & [0.07, 0.03]& (4)\\
				G15.03-0.67\tablefootmark{{\rm ({\it b})}}  &2.0&6.4&19.2  &32.8$\pm$0.7 &258.2&0.25&23.32&[5.43, 6.13]&  --  & --  & --                   &--          &--          &18.6$^{+0.1}_{-0.1}$ &3.7$^{+0.1}_{-0.1}$ &13.2$^{+0.1}_{-0.1}$&[14.0, 13.4]&[3.4, 0.8] & --   & --  & --                   &          -- & --          & (4)\\
				G335.79+0.17                                &3.3&5.5&-50.5 &23.1$\pm$2.9 & 34.4&0.17&23.26&[5.54, 6.24]&  --  & --  & --                   &--          &--          &-50.2$^{+0.1}_{-0.1}$&4.5$^{+0.1}_{-0.1}$ &13.4$^{+0.1}_{-0.1}$&[14.3, 13.7]&[5.3, 1.1] & -50.3$^{+0.4}_{-0.6}$ & 3.0$^{+0.9}_{-0.8}$ & 11.3$^{+0.1}_{-0.2}$ &[11.4, 11.2] & [0.02, 0.01]& (4)\\
				G337.92-0.48                                &2.9&5.8&-40.6 &34.6$\pm$0.9 &492.0&0.09&23.47&[6.03, 6.73]&  --  & --  & --                   &--          &--          &-40.8$^{+0.1}_{-0.1}$&4.8$^{+0.2}_{-0.2}$ &12.8$^{+0.1}_{-0.1}$&[13.1, 12.8]&[0.3, 0.1] & --   & --  & --                   &          -- & --          & (4)\\
				G351.13+0.77                                &1.3&7.1&-5.6  &19.1$\pm$0.3 &  6.3&0.15&22.60&[4.94, 5.64]&  --  & --  & --                   &--          &--          &-5.6$^{+0.1}_{-0.1}$ &1.4$^{+0.2}_{-0.2}$ &12.3$^{+0.1}_{-0.1}$&[13.8, 13.1]&[3.1, 0.6] & -5.7$^{+0.1}_{-0.1}$ & 1.1$^{+0.3}_{-0.2}$ & 11.4$^{+0.1}_{-0.1}$ &[12.0, 11.5] & [0.16, 0.06]& (4)\\
				G351.16+0.70\tablefootmark{{\rm ({\it b})}} &1.3&7.1& -6.4 &21.9$\pm$3.4 &  7.5&0.09&23.67&[6.23, 6.93]&  --  & --  & --                   &--          &--          &-6.4$^{+0.1}_{-0.1}$ &4.4$^{+0.1}_{-0.1}$ &13.1$^{+0.1}_{-0.1}$&[13.4, 13.2]&[0.7, 0.3] & --   & --  & --                   &          -- & --          & (4)\\
				G351.25+0.67                                &1.3&7.1& -1.8 &31.3$\pm$6.7 &  0.4&0.13&23.44&[5.84, 6.54]&  --  & --  & --                   &--          &--          &-2.4$^{+0.1}_{-0.1}$ &3.1$^{+0.2}_{-0.3}$ &12.4$^{+0.1}_{-0.1}$&[12.8, 12.4]&[0.2, 0.1] & --   & --  & --                   &          -- & --          & (4)\\
				\multirow{2}{*}{G353.41-0.36\tablefootmark{$\dagger$, $\ddagger$, {\rm ({\it b})}}}&3.1&5.3&-16.2&28.3$\pm$0.2&3.2&0.25&23.49&[5.61, 6.31]&--&--&--&--&--&-19.4$^{+0.2}_{-0.4}$&2.6$^{+0.7}_{-0.4}$& (c1) 11.8$^{+0.1}_{-0.1}$& [13.2, 12.6]& [0.4, 0.2]& -15.3$^{+0.4}_{-0.2}$ & 1.3$^{+0.4}_{-0.4}$ & (c1) 11.6$^{+0.2}_{-0.1}$ &[11.7, 11.5] & [0.06, 0.02]& \multirow{2}{*}{(4)}\\[-2pt]
				&  & & & & & & & & & & & & &-14.6$^{+0.1}_{-0.1}$&4.1$^{+0.2}_{-0.3}$& (c2) 12.5$^{+0.1}_{-0.2}$ &[13.8, 13.2] & [1.4, 0.4] & -14.5$^{+0.1}_{-0.2}$ & 1.2$^{+0.3}_{-0.4}$ & (c2) 11.2$^{+0.1}_{-0.2}$ & [11.3, 11.2] & [0.03, 0.01] & &\\
				
				\hline                                                                                                                       
			\end{tabular}
			\tablefoot{Physical properties taken from \citet{Urquhart22}. The sources are separated, from top to bottom, by their evolutionary class: (1) Quiescent; (2) Protostellar; (3) YSO and (4) PDR. The reference velocities of the individual sources ($V_{\rm 0}$)  were derived using different chemical tracers. The error associated to $N$(H$_2$) is 20\% for each source, while those on $T_{\rm dust}$ are shown in parentheses; \tablefoottext{a}{The lower limit of the volume density range was derived considering the clump size and H$_2$ column density, i.e. $n$(H$_2$) = $N$(H$_2$)/2\:$R_{\rm eff}$, while the upper limits incorporate a factor of 5 uncertainty in the $R_{\rm eff}$.}; \tablefoottext{b}{$T_{\rm dust}$, $N$(H$_2$) and from \citet{Giannetti17_june}}. The symbols ($\dagger$) and ($\ddagger$) identify sources that exhibit multiple spectral components in \nnhp and \nndp, respectively. These are labelled as ``(c1)'' and ``(c2)'' in the column listing $N^{\rm LTE}$. }
		\end{table}
	\end{landscape}
	
	\begin{figure*}
		\centering
		\includegraphics[width=0.87\hsize]{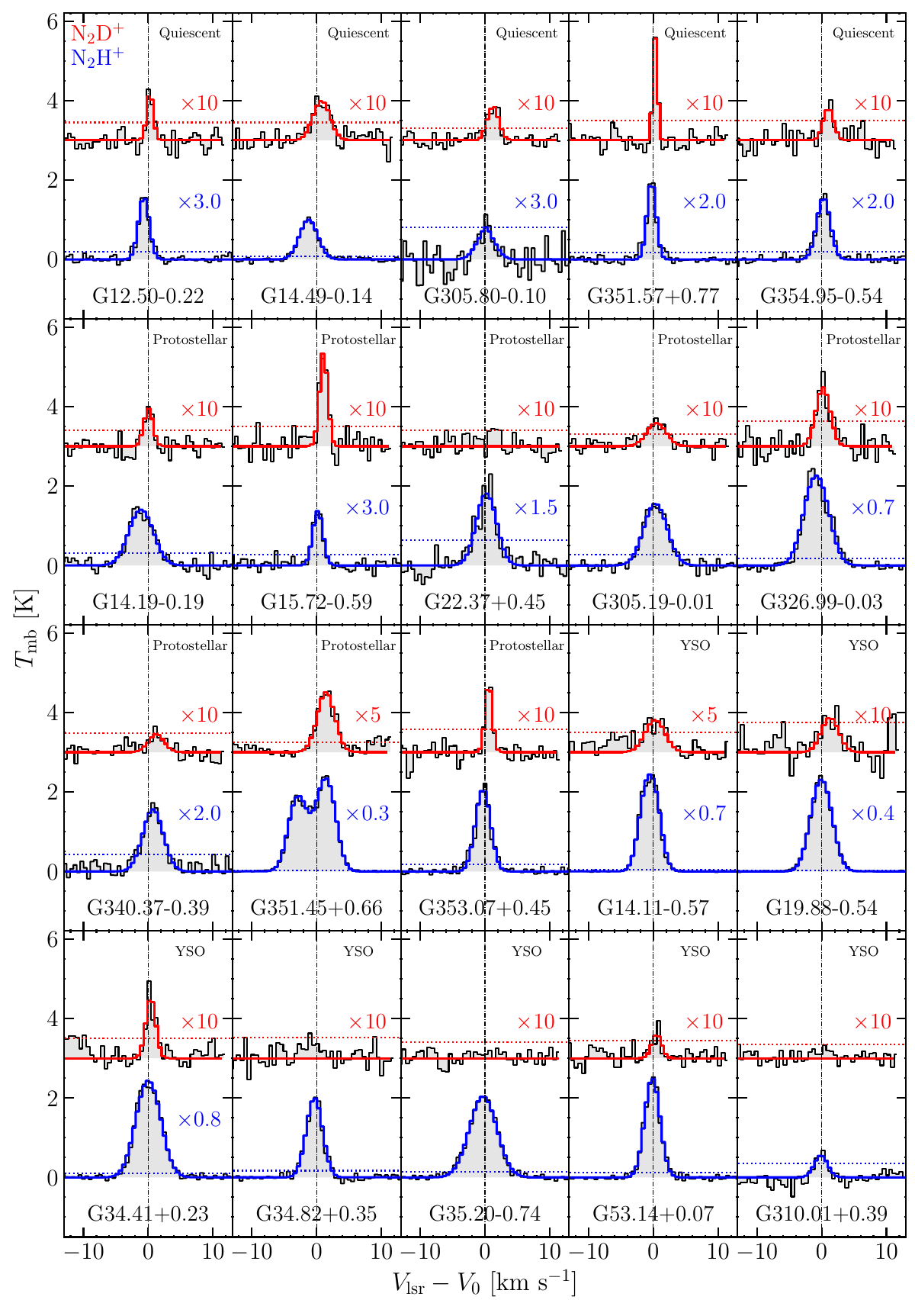}
		\caption{Spectral overview of the N$_2$D$^+$ (top) and N$_2$H$^+$ (bottom) lines observed in the entire sample. In each panel, the source name and evolutionary stage are shown in the lower centre and upper right corners, respectively. The \texttt{MCWeeds} models are shown in red for N$_2$D$^+$ and in blue for N$_2$H$^+$. The dotted lines represent the 3$\sigma$ noise levels. The vertical dashed lines indicate the $V_{\rm lsr}$ of the sources (see Table~\ref{tab:summary}). Spectra have been multiplied by an arbitrary factor, shown in each panel.}\label{fig:spectra}%
	\end{figure*}
	\setcounter{figure}{0}
	\begin{figure*}
		\centering
		\includegraphics[width=0.87\hsize]
		{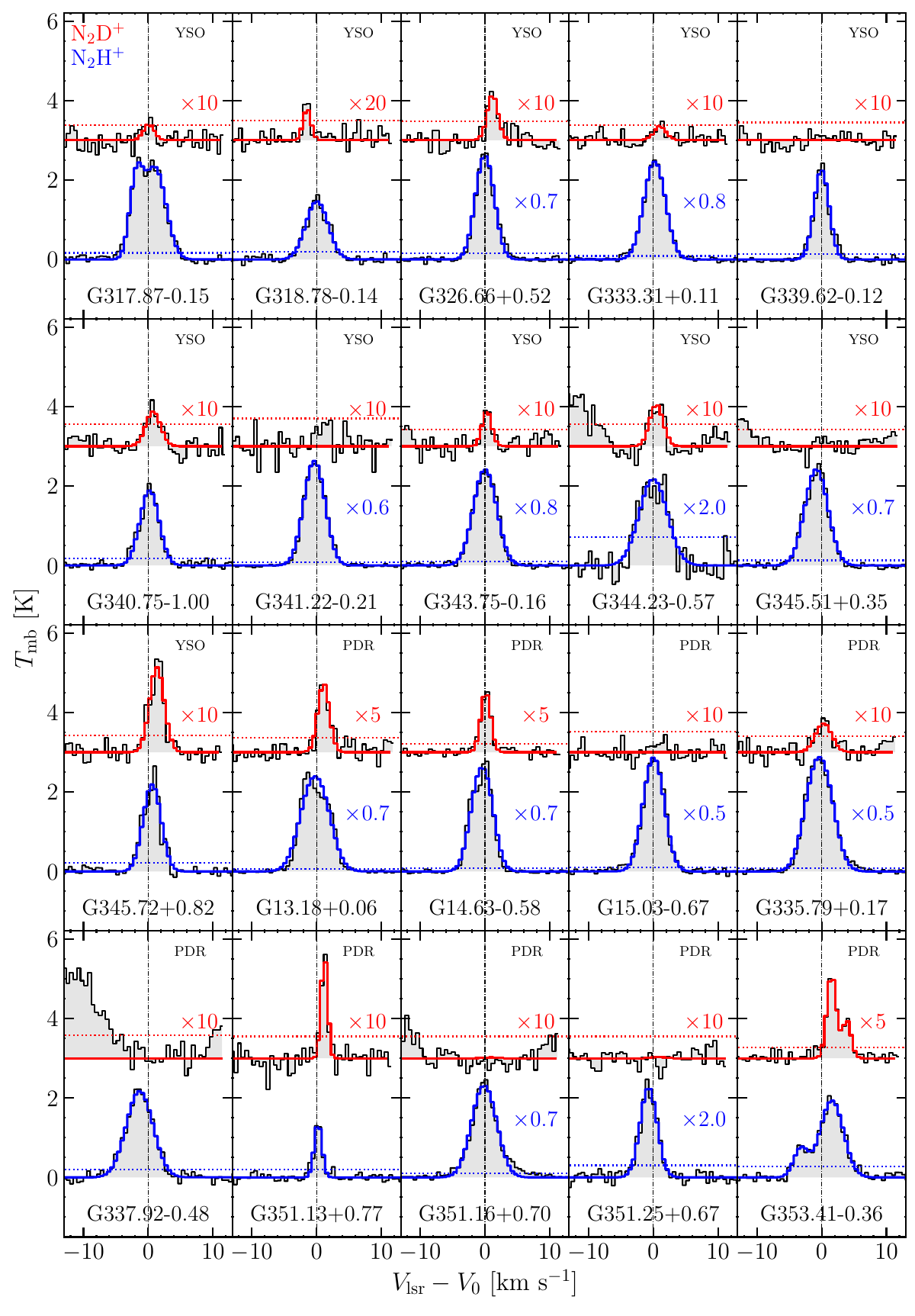}
		\caption{Continued.}%
	\end{figure*}
	
\end{appendix}

\end{document}